\documentclass[useAMS,usenatbib]{mn2e}
\usepackage{epsfig}
\usepackage{times}
\usepackage{subfigure}
\newcommand{\qo}{\"{o}}

\title[High Resolution Optical Spectroscopy of IRAS
18095+2704]{High Resolution Optical Spectroscopy of the F Supergiant Proto-Planetary Nebula V887 Her=IRAS 18095+2704}

\author[T. \c{S}ahin, D. L. Lambert, V. G. Klochkova, and N.S. Tavolganskaya]
{T. \c{S}ahin\thanks{E-mail:[sahin, dll]@astro.as.utexas.edu}$^{1}$, David L. Lambert$^{1}$, V. G.
Klochkova$^{2}$, and N.S. Tavolganskaya$^{2}$\\
$^{1}$Department of Astronomy and The W.J.  McDonald Observatory, University of Texas, Austin, TX 78712, USA\\
$^{2}$Special Astrophysical Observatory, Nizhnij Arkhyz Stavropol Territory, Karachai-Cherkessia, 369167 Russia}

\begin{document}

\date{Accepted .
  Received ;}

\pagerange{\pageref{firstpage}--\pageref{lastpage}} \pubyear{2010}

\maketitle

\label{firstpage}

\begin{abstract}

An abundance analysis is presented for IRAS 18095+2704 (V887 Her), a post-AGB star and proto-planetary nebula. The
analysis is based on high-resolution optical spectra from the McDonald Observatory and the Special Astrophysical
Observatory. Standard analysis using a classical Kurucz model atmosphere and the line analysis program MOOG provides
the atmospheric parameters: $T_{\rm eff} = 6500$ K, $\log g = +0.5$, and a microturbulent velocity $\xi = 4.7$ km
s$^{-1}$ and [Fe/H] $= -0.9$. Extraction of these parameters is based on excitation of Fe\,{\sc i} lines, ionization
equilibrium between neutral and ions of Mg, Ca, Ti, Cr, and Fe, and the wings of hydrogen Paschen lines.  Elemental
abundances are obtained for 22 elements and upper limits for an additional four elements. These results
show that the star's atmosphere has not experienced a significant number of C- and $s$-process enriching thermal
pulses. Abundance anomalies as judged relative to the compositions of unevolved and less-evolved normal stars of a
similar metallicity include Al, Y, and Zr deficiencies with respect to Fe of about 0.5 dex. Judged by composition,
the star resembles a RV Tauri variable  that has been mildly affected by dust-gas separation reducing the abundances
of the elements of highest condensation temperature. This separation may occur in the stellar wind. There are indications that
the standard 1D LTE analysis is not entirely appropriate for IRAS 18095+2704. These include a supersonic macroturbulent velocity of 23 km s$^{-1}$, emission in H$\alpha$ and the
failure of predicted profiles to fit observed profiles of H$\beta$ and H$\gamma$.

\end{abstract}


\section{Introduction}

\noindent Stars of low and intermediate mass with initial masses between 0.8$M_\odot$ -- 8$M_\odot$ evolve to the
asymptotic giant branch (AGB). Then, thanks to severe mass loss, the AGB star evolves rapidly at nearly constant
luminosity to higher effective temperatures to the white dwarf cooling track. Typical stellar lifetimes of post-AGB
stars are expected to be of the order of 10$^{\rm 4}$ years (Sch\qo nberner 1983). The gas lost by the AGB star forms
a circumstellar shell. When the post-AGB star is cool, the dust in the shell heated by stellar radiation provides an
infrared excess. When the star has traversed the top of the H-R diagram to higher effective temperatures, the
circumstellar gas is ionized. Then, the star is said to have evolved to reached the  proto-planetary nebula stage.
Shortly after this, the post-AGB star has evolved to become a planetary nebula with a hot white dwarf as the central
star. Determinations of the chemical composition for post-AGB star hold the potential of yielding insights into the
chemical history of the AGB star and its conversion by mass loss to its slimmer post-AGB form.
\\
\noindent  In this paper, we present a determination of the chemical composition of IRAS
18095+2704, a post-AGB star with a substantial dusty circumstellar shell. The discovery of the
optical counterpart IRAS 18095+2704 was made by Hrivnak, Kwok, \& Volk (1987, 1988). This $V$ =
10.4 mag star is a high-latitude F supergiant with a large far-IR excess. In the {\it Catalog of
Low Resolution IRAS Spectra} assembled by Hrivnak, Kwok, \& Volk (1988), the star has a peculiar
IR continuum slope at wavelengths shortward of the 10 \micron\, silicate emission feature.
According to Volk \& Kwok (1987), this peculiar continuum shape is a result of a detached dust
shell. Observational evidence for an expanding shell came from Lewis, Eder, \& Terzian
(1985) and Eder, Lewis, \& Terzian (1988) via detection of OH maser emission at 1612 and 1665/67 MHz
from the Arecibo telescope. Gledhill et al. (2001) from imaging polarimetry report an extended
envelope or a reflection nebula around the star.
\vskip 0.2 cm
 \begin{figure*}
 \centering
 \includegraphics[width=18.5cm,height=10.cm,angle=0]{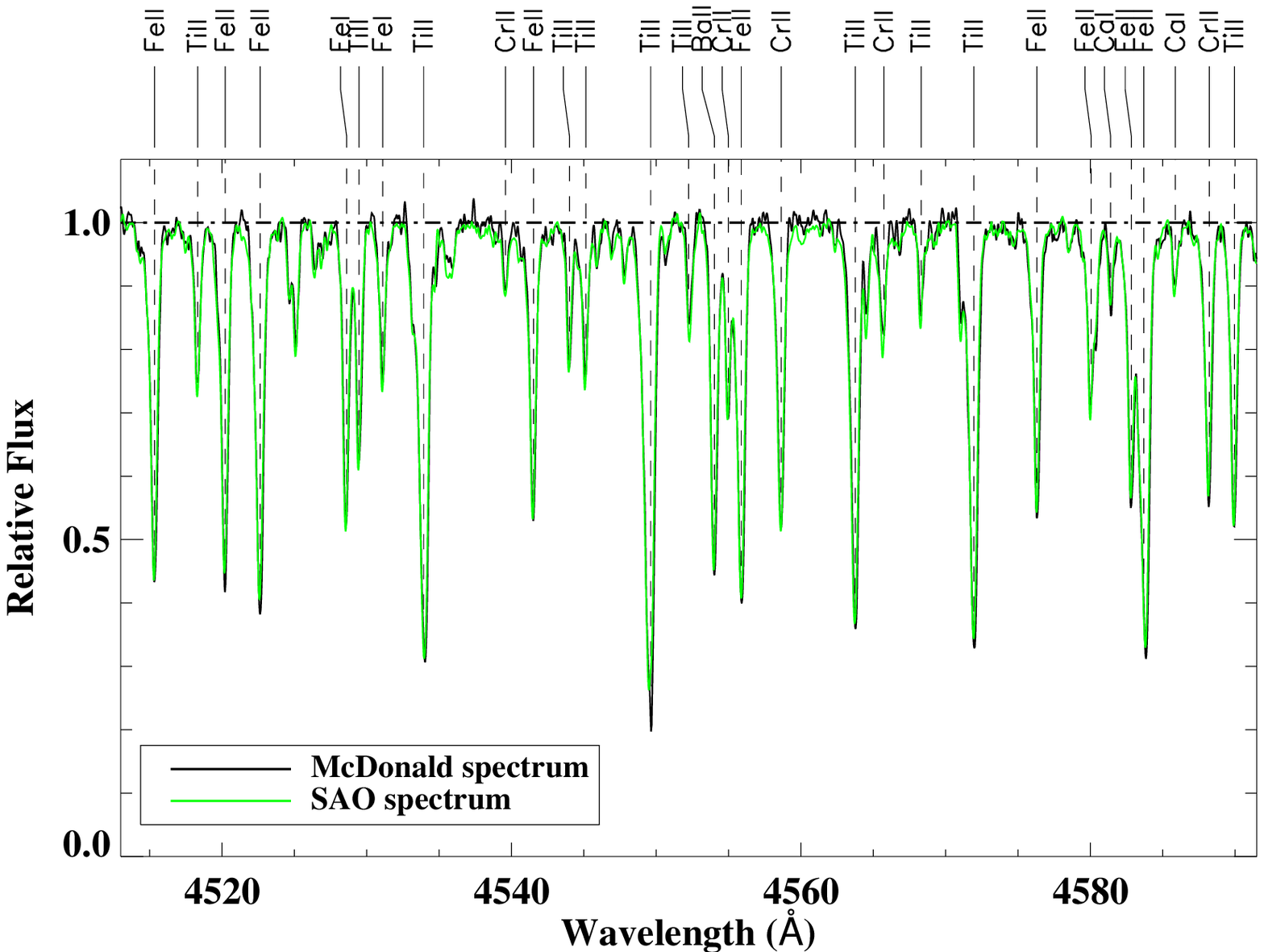}
   \caption{The spectrum for IRAS 18095+2704 over the wavelength region  4513-4592 \AA\ . The McDonald spectrum is plotted in black and the SAO spectrum in green. 
Selected lines are identified.}
      \label{f_spectrum_part}
 \end{figure*}

\noindent The pioneering study of IRAS 18095+2704's composition was reported by Klochkova (1995) from echelle spectra ($R
= 24\,000$) covering the wavelength ranges 5050 \AA\ to 7200 \AA\ and 5550 \AA\ to 8700 \AA. Abundances of 26 elements
were obtained. The star was found to be moderately metal poor, [Fe/H]= $-$0.78.\footnote{Standard notation is used for
quantities [X] where [X]=$\log$(X)$_{\rm star}-\log$(X)$_\odot$.} with a relative enrichment of C and N, i.e., [C/Fe]=0.5
and  [N/Fe]=0.5, as might be expected of a post-AGB star evolved from a C-rich AGB star. Although other elements up
through the Fe-peak had roughly their anticipated abundances, two results drew comment. First, there was a difference in
abundances [X/H] derived from neutral and first-ionized lines of several elements, i.e., differences of 1.6 (Ti), 1.4 (V),
1.0 (Cr), and 2.2 dex (Y). For Fe, this difference was zero because it was the condition enforced in determining the
surface gravity. Second, the accessible lanthanides (La, Pr, Nd and Eu) represented by ionized lines were overabundant by
about [X/Fe] $\simeq +0.7$ relative to what is expected for an unevolved metal-poor star. Although one might attribute
this overabundance to $s$-process enrichment expected of a post-AGB star, one notes that Eu, predominantly an $r$-process
element, had the highest overabundance ([Eu/Fe] = +1.2), Ba was not overabundant ([Ba/Fe] = $-$0.2), and yttrium, the sole
representative of lighter $s$-process species, as analyzed from Y\,{\sc ii} lines, was also not overabundant ([Y/Fe]
$\simeq 0$).
\vskip 0.2 cm
\noindent In this paper, we determine afresh the composition of {\sc IRAS18095+2704} from echelle optical spectra: one
spectrum was obtained at the W.J. McDonald Observatory and another at the Special Astrophysical Observatory (SAO). In
addition to examining the unusual results noted above, we seek an interpretation of the star's the composition in light
of its proposed status as a slightly metal-poor post-AGB star.
\vskip 0.2 cm
 \begin{figure}
 \centering
 \includegraphics[width=0.99\columnwidth,height=80mm,angle=0]{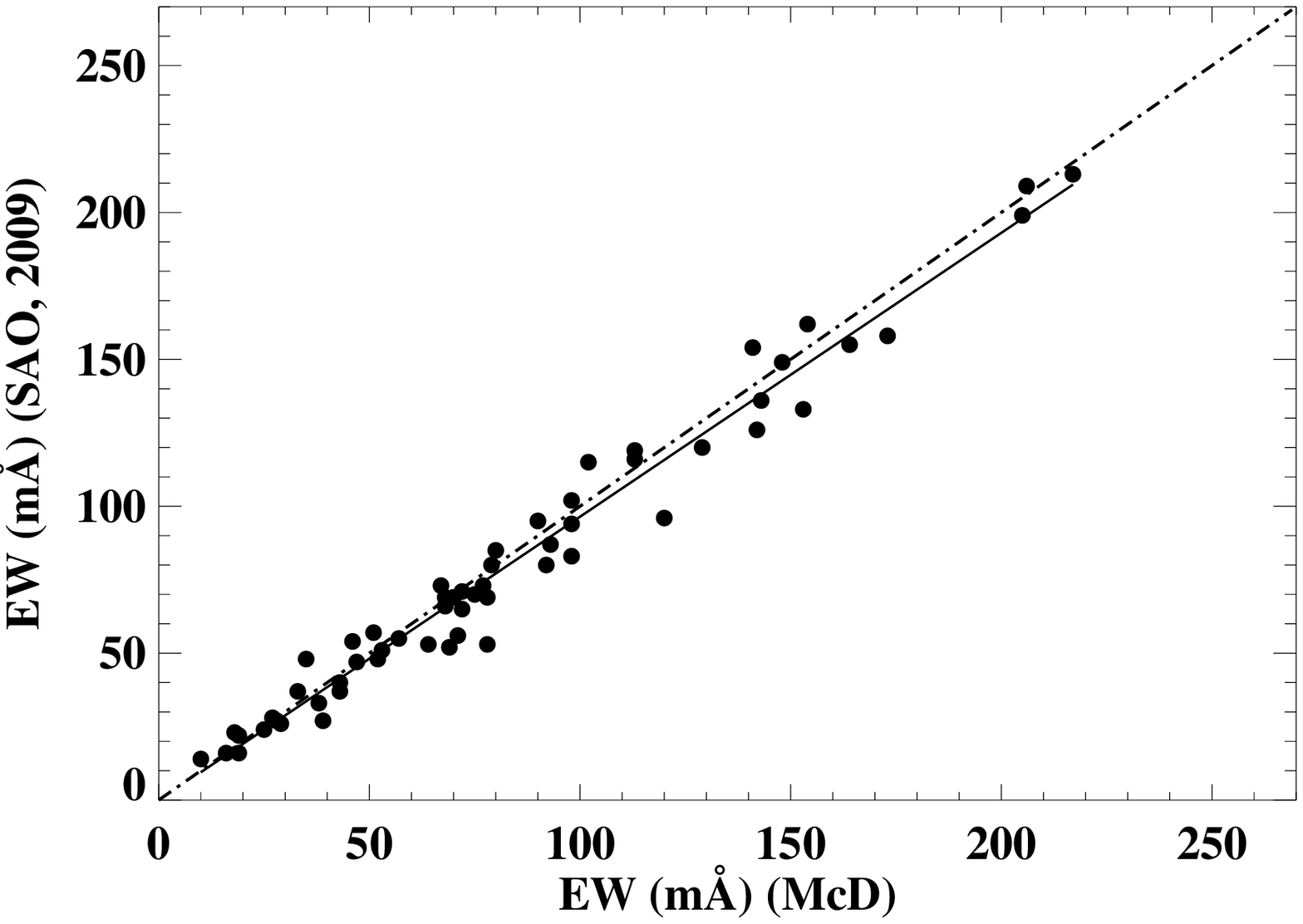}
\caption{Equivalent widths (EW) for 69 lines in the McD 
spectrum versus the EWs of the same lines from SAO (2009) spectrum. The solid line is the
least-squares fit. The dash-dot line indicates EW quality.}
      \label{f_ew_comparison}
 \end{figure}

\section[]{Observations and Data Reduction}

\noindent The McDonald spectrum for the abundance analysis was obtained on the night of 2008 August 10
(JD 2454688.7)
at the 2.7 meter Harlan J. Smith reflector with the Tull cross-dispersed \'{e}chelle
spectrograph (Tull et al. 1995) at a spectral resolution of $\lambda/d\lambda \simeq 60,000$. The spectrum covers the wavelength ranges 3800 \AA\ to 10\,500 \AA\ with no gaps in the
wavelength ranges 3800 \AA\ to 4885 \AA\ and 5020 \AA\ to 5685 \AA\  but coverage is incomplete but
substantial  beyond 5700 \AA; the effective short and long wavelength limits are set by
the useful  S/N ratio. A ThAr hollow cathode lamp provided the wavelength calibration. Flat-field
and bias exposures completed the calibration files. The signal-to-noise ratio ranges between 40
and 90 per pixel, not only changing with the blaze function within echelle
orders but also star brightness between echelle
orders.

\begin{table}
    \caption[]{Absorption components of Na\,{\sc i} D$_{\rm 1}$ (5895.923 \AA\ ) and Na\,{\sc i} D$_{\rm 2}$ (5889.953
    \AA\ ) lines in the McD spectrum of IRAS 18095+2704 (V887 Her). V$_{\odot}$ are the heliocentric radial velocities of
    the components.}
       \label{}
   $$
       \begin{array}{@{}c|r|r@{}}
          \hline
          \hline
            &  NaD1       &  NaD2        \\
\cline{2-3}
 $Component$ &  V_{\odot}    & V_{\odot} \\
\cline{2-3}
            &  (km s^{-1})  & (km s^{-1}) \\
\cline{1-3}
     $1$    & -43.6      &  -41.7      \\
     $2$    & -27.1      &  -26.1      \\
     $3$    & -18.1      &  -17.0      \\
     $4$    & -6.8       &   -4.5      \\
          \hline
\hline
       \end{array}
   $$
\end{table}

\noindent The McD observations were reduced using the {\sc STARLINK} echelle reduction package
{\sc ECHOMOP} (Mills \& Webb 1994). The spectra were extracted using {\sc ECHOMOP'S}
implementation of the optimal extraction algorithm developed by Horne (1986). {\sc ECHOMOP}
propagates error information based on photon statistics and readout noise throughout the
extraction process. The bias level in the overscan area was modeled with a polynomial and
subtracted. The scattered light was modeled and removed from the spectrum. In order to correct
for pixel-to-pixel sensitivity variations, `flatfield' exposures from a halogen lamp were
used. Individual orders were cosmic-ray cleaned, and continuum normalized with bespoke echelle
reduction software in {\sc IDL} (\c{S}ahin 2008). Reduced spectra were transferred to the {\sc
STARLINK} spectrum analysis program {\sc DIPSO} (Howarth et al. 1998) for further analysis
(e.g. for equivalent width measurement). In equivalent width measurements, local continua on
both side of the lines were fitted with a first-degree polynomial then equivalent widths were
measured with respect to these local continua using a fitted Gaussian profile. For strong
lines, a direct integration was preferred to the Gaussian approximation. The errors for each
equivalent width measurement were determined on the basis of scatter of linear continuum fit
and signal-to-noise ratio of each measured line in the spectra. Errors on the measured
equivalent widths are calculated using the prescriptions given by Howarth \& Phillips (1986).
\vskip 0.2 cm
\noindent The SAO spectrum was obtained on the night of 2009 June 10 (JD 2452993.4) by
VK and NST with the NES echelle spectrograph
mounted at the Nasmyth focus of the 6-m telescope of the Special Astrophysical
Observatory (Panchuk et al. 2007) with  a
2048 X 2048 CCD with an image slicer (Panchuk et al. 2007)
and a spectral resolution of $\lambda/d\lambda \ge 60,000$. A
modified ECHELLE context (Yushkin \& Klochkova 2005) of {\sc MIDAS} package
was used to extract one-dimensional vectors from the
two-dimensional echelle spectra.
Wavelength calibration was performed using a hollow-cathode Th-Ar lamp. The wavelength
coverage for the SAO spectrum was 4460 -- 5920 \AA\ .
Measurement of equivalent widths was carried out as for the McDonald spectrum.
\vskip 0.2 cm
 \begin{figure*}
 \centering
 \includegraphics[width=184mm,height=86mm,angle=0]{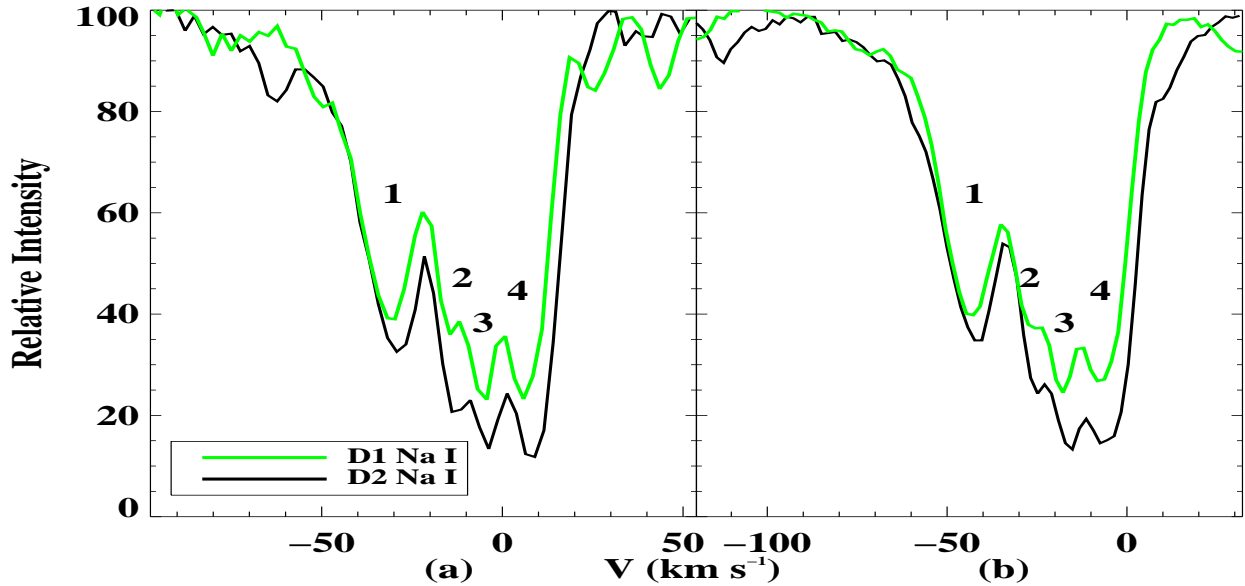}
\caption{The profile of the D-lines of Na\,I  a) in the McDonald 2008  spectrum  and b) in the SAO 2009
spectrum of IRAS 18095+2704.
	 Telluric H$_2$O lines have not been 
removed.}
      \label{NaD_lines}
 \end{figure*}

The agreement between the McDonald and SAO spectra is satisfactory, i.e., the two spectra are similar as to line width, depth and
equivalent width for weak to strong lines. This is shown by
the section
of the reduced spectra illustrated in Figure~\ref{f_spectrum_part}.
A sample of apparently unblended lines was selected from across the
common wavelength interval and their equivalent widths (EWs)
measured in both the McDonald and the SAO spectra. The comparison
of EWs shown in Figure~\ref{f_ew_comparison} shows good agreement between
the two sets of measurements. Across the common wavelength interval, we
compare EWs from McDonald and SAO spectra, especially for lines at the
limit of detection and for elements represented by just one or two lines. The SAO spectrum was used to provide lines
that fell in the inter-order gaps of the McDonald spectrum.

\section[]{General features of the spectra}

\noindent Although the data are sparse, the
star is probably not a large amplitude velocity variable. Klochkova (1995) reported a heliocentric radial velocity of
$-$32.5$\pm$0.4 km s$^{-1}$. Hrivnak, Kwok, \& Volk (1988) report $-$30$\pm$2 km s$^{-1}$ from an
unspecified number of measurements but add that there is `an indication of variability'. The McD
spectrum gives $-$30$\pm$1 km s$^{-1}$ from the metal lines. The 2009 SAO  spectrum gives
$-$31.8$\pm$1.7 km s$^{-1}$. Although Lewis, Eder, \& Terzian (1985) and Eder, Lewis, \& Terzian
(1988) cite the heliocentric velocity as $-$17.4 km s$^{-1}$ from their observed OH radio lines,
Hrivnak (2009, private communication)  indicates that this value resulted from an incorrect
conversion of LSR to heliocentric velocity and a velocity of about $-30$ km s$^{-1}$ is obtained
from the OH velocities.
\vskip 0.2 cm

\noindent The Na D lines show four components. Figure~\ref{NaD_lines}
shows the Na D from the McDonald and SAO spectra.  Heliocentric velocities of the four
principal components in the McDonald spectrum are listed in Table 1. The SAO spectrum gives
similar velocities.
Stellar photospheric Na lines at about $-$30 km s$^{-1}$
must be largely masked by these multiple circumstellar components. Component 1, if not
an interstellar component, represents an outflow at a velocity of about 12 km s$^{-1}$.
Components 3 and 4 are falling toward the star at velocities
of about 12 and 25 km s$^{-1}$, respectively.
\vskip 0.2 cm
 \begin{figure}
 \centering
 \includegraphics[width=0.99\columnwidth,height=85mm,angle=0]{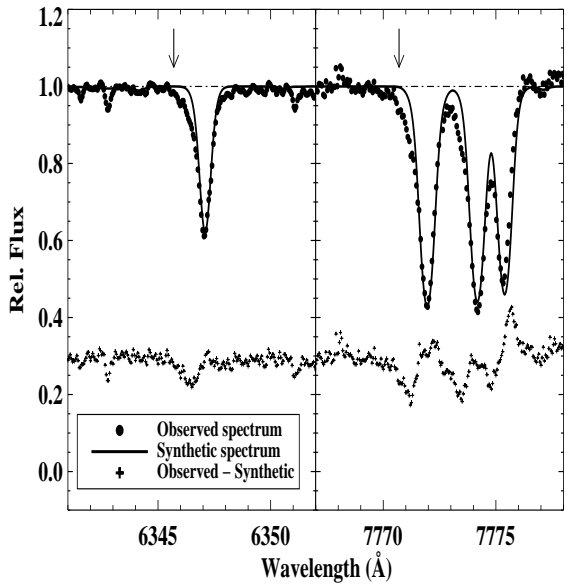}
\caption{Profiles of Si\,{\sc ii} 6347 \AA\ line and O\,{\sc i} 7772 \AA\ triplet in the McDonald
spectrum. The best-fit synthetic spectra (in green) fail to account for the extended blue
wings. The bottom of each panel shows the residuals shifted by 0.3 relative flux for clarity.}
      \label{f_wind}
 \end{figure}

\noindent Weak emission in the blue and red wings of the H$\alpha$ profile flanking a deep narrow
absorption core was reported by Klochkova (1995 - see also Tamura, Takeuti \& Zalewski 1993). On the
McDonald spectrum, H$\alpha$ occurs at the very edge of an order but a very shallow a deep core
flanked by red emission is seen. H$\beta$ and higher lines in the Balmer series and Paschen lines are purely in absorption. The 2009 SAO spectrum did not include
H$\alpha$.
\vskip 0.2 cm

\noindent  The stellar absorption lines are broad.
If one accepts classical notions of microturbulence and macroturbulence, this
width suggests substantial macroturbulence in the atmosphere. The
microturbulence is about 5 km s$^{-1}$ (see below). The instrumental width
is about 5 km s$^{-1}$. Correcting for the microturbulence and the instrumental
width, the macroturbulence is estimated to be about 23 km s$^{-1}$.
This represents a highly supersonic velocity.
\vskip 0.2 cm
 \begin{figure}
 \centering
 \includegraphics[width=0.99\columnwidth,height=80mm,angle=0]{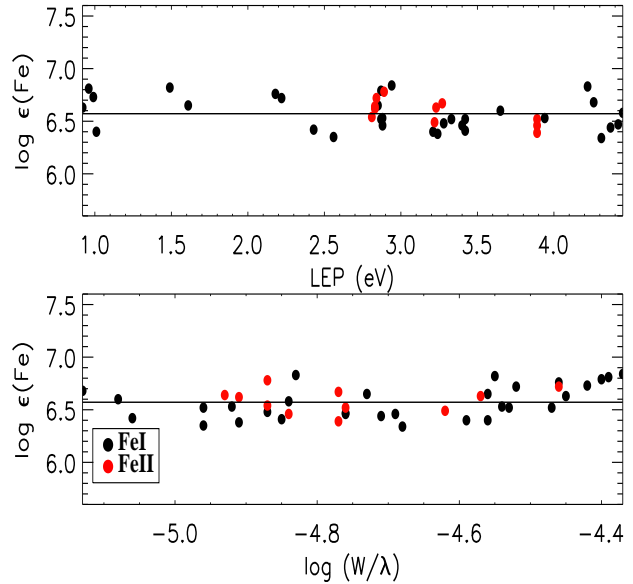}
\caption{Determination of atmospheric parameters $T_{\rm eff}$ and $\xi$ using abundance ($\log \epsilon(Fe)$) as
a function of both lower level excitation potential
(LEP) and reduced equivalent width ($\log (W/ \lambda)$). In all the panels, the
solid black line is the least-square fit to the data. The zero slope of this line is for $T_{\rm eff}$=6500 K. The ionized (in red) iron lines were over plotted.}
     \label{f_excitation}
 \end{figure}

\noindent Strong lines clearly show asymmetric profiles with an extended blue wing. This is illustrated in
Figure~\ref{f_wind} where the O\,{\sc i} triplet lines at 7771-7775\AA\ and the Si\,{\sc ii} 6347\AA\ line are shown.
The blue wing extension extends from about $-$25 km s$^{-1}$ to $-$50 km s$^{-1}$ with respect to the photospheric
velocity. Inspection of strong unblended lines shows that the blue asymmetry is present also for low excitation strong
lines (e.g., Mg\,{\sc i}b 5167 \AA\ , 5172 \AA\ , and Sr\,{\sc ii} 4215 \AA\ ). This outward motion may represent the early stages of a stellar wind or
inhomogeneities (i.e., stellar super-granulation) in the photosphere.
\vskip 0.2 cm

 \begin{figure}
 \centering
 \includegraphics[width=0.99\columnwidth,height=80mm,angle=0]{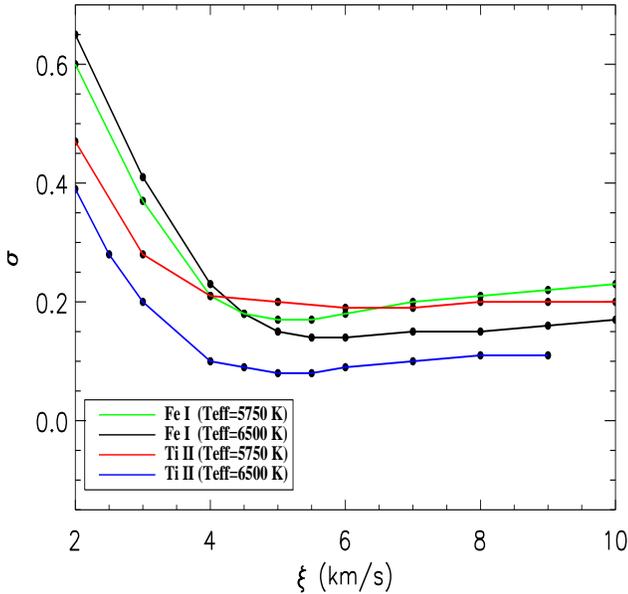}
\caption{The standard deviation of the Fe and Ti abundances from the suite of Fe\,{\sc i} and Ti\,{\sc ii}
lines as a function of the microturbulence $\xi$.}
     \label{f_micro}
 \end{figure}

\section[]{ABUNDANCE ANALYSIS -- The Model Atmospheres and Stellar Parameters}

\noindent The abundance analysis was performed using the local thermodynamic equilibrium (LTE)
stellar line analysis program {\sc MOOG} (Sneden 2002). Model atmospheres were obtained by
interpolating in the ATLAS9 model atmosphere grid (Kurucz 1993). The models are line-blanketed
plane-parallel uniform atmospheres in LTE and hydrostatic equilibrium with flux (radiative plus
convective) conservation. A model is defined by an effective temperature $T_{\rm eff}$, surface
gravity $g$, chemical composition as represented by metallicity [Fe/H] and a microturbulence
velocity $\xi$ of 2 km s$^{-1}$.
\vskip 0.2 cm
\begin{figure}
 \centering
 \includegraphics[width=88mm,height=85mm,angle=0]{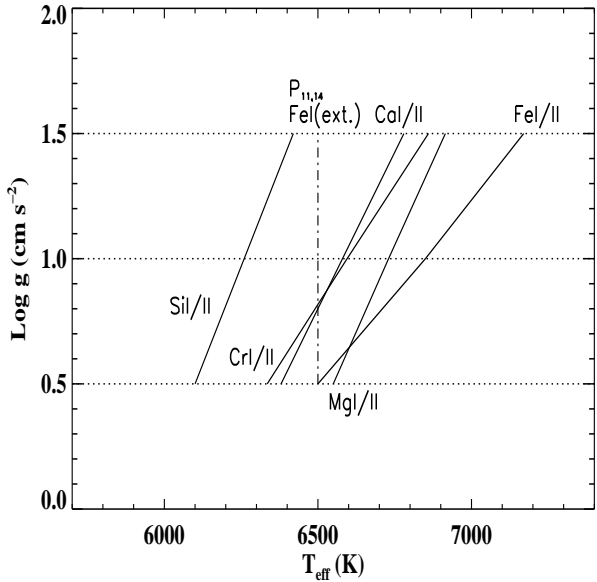}
\caption{The $T_{\rm eff}$ versus $\log g$ plane showing the various loci discussed in the text. 
Loci include those derived from the Paschen line profiles (P$_{\rm 11,14}$, dashed line), the
excitation of the Fe\,{\sc i} lines (Fe\,{\sc i}(ext), dashed line), and the imposition of
ionization equilibrium for Mg, Si, Ca, Cr, and Fe.}
      \label{f_teff_logg}
 \end{figure}

\noindent  Several of the assumptions adopted in the construction and application of the
model atmospheres are of uncertain validity when considering post-AGB stars and
supergiants in general. The presence of supersonic levels of macroturbulence and hints of
a stellar wind seem incompatible with the assumption of
hydrostatic equilibrium. Departures from LTE are
 probable for these low density atmospheres.
The real atmosphere is likely to depart from the uniform
plane-parallel layered theoretical construction.
These qualifying remarks should be borne in mind when interpreting the derived abundances.
\vskip 0.2 cm

 \begin{figure*}
 \centering
 \includegraphics[width=183mm,height=95mm,angle=0]{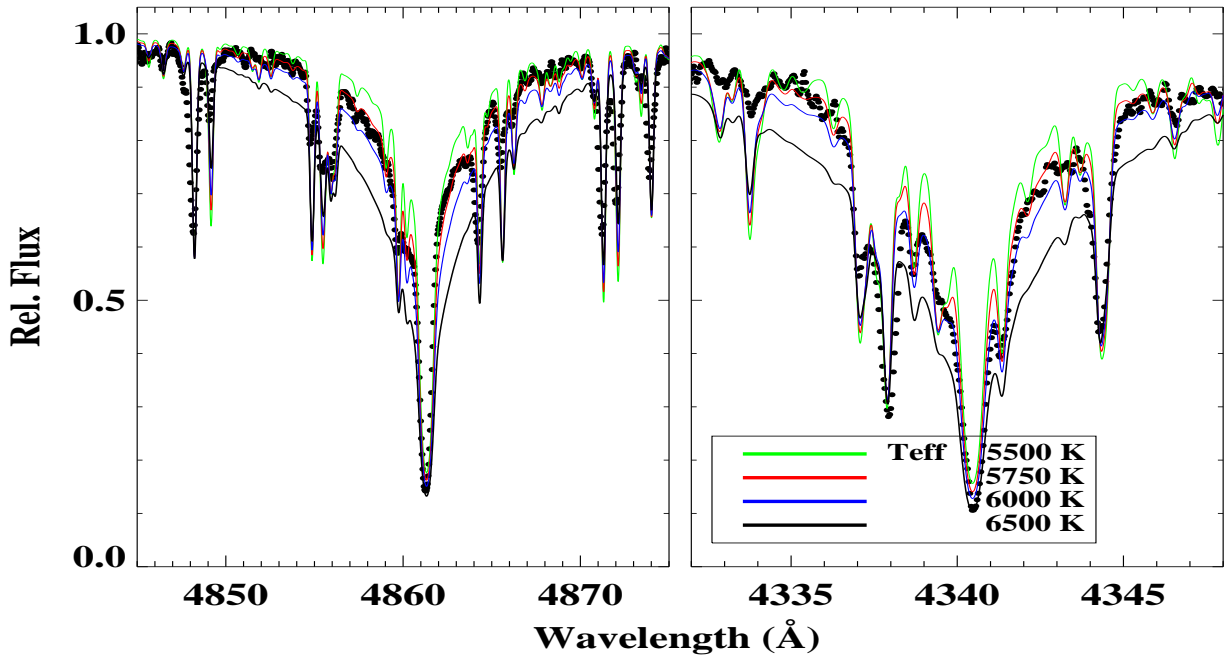}
\caption{Observed and model line profiles for H${\beta}$ and H${\gamma}$.  The green, red,
blue, and black lines show the theoretical profiles for $T_{\rm eff}$ = 5500 K, 5750 K, 6000 K, and 6500
K, respectively, all for a surface gravity $\log$ g = 0.5 dex.}
     \label{f_h_beta}
 \end{figure*}

 \begin{figure*}
 \centering
 \includegraphics[width=180mm,height=115mm,angle=0]{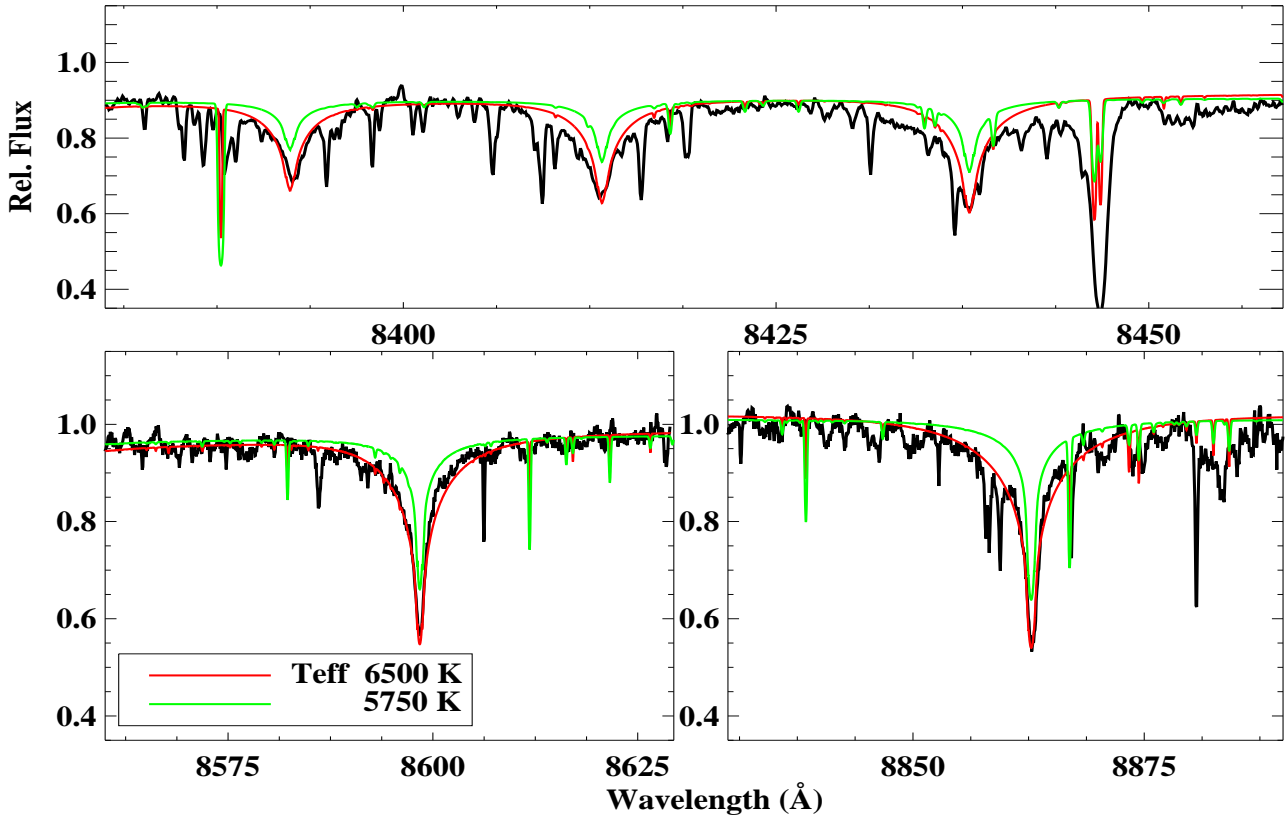}
\caption{Observed and model line profiles for hydrogen Paschen lines. 
The green and red lines show the theoretical profiles for $T_{\rm eff}$ =  5750 K, and 6500 K, respectively, both
 for a surface gravity $\log$ g = 0.5 dex.
}
     \label{f_paschen}
 \end{figure*}

\subsection{Spectroscopy - Fe\,{\sc i} and Fe\,{\sc ii} lines}

\noindent A standard spectroscopic method of
determining the effective temperature, surface gravity, and the microturbulence
using Fe\,{\sc i} and Fe\,{\sc ii} was applied. Application of strict
criteria\footnote{Lines that are well defined and not blended.} in the selection of suitable lines provided a list of
31 Fe\,{\sc i} lines with lower
excitation potentials (LEP) ranging from 0.9 to 4.5 eV and EWs of up to 224 m\AA\ and
11 Fe\,{\sc ii}  lines with excitation potentials of 2.8 to 3.9 eV and EWs of up to
154 m\AA.
 Only six
Fe\,{\sc i} lines have EW greater than 150 m\AA.
These lines are listed in Table 2. The $gf$-values are taken from F\"{u}hr \& Wiese
(2006).
\vskip 0.2 cm

\noindent One estimate of the temperature is found from Fe\,{\sc i} lines by excitation balance.
The value for $T_{\rm eff}$  is chosen so that the abundance is independent of a line's lower
level excitation potentials (LEP). The microturbulence is determined by adjusting the $\xi$ so
that the abundances were independent of reduced equivalent width ($W/\lambda$). For our
sample of Fe\,{\sc i} lines, these two conditions are imposed simultaneously (see Figure 5). The
microturbulence may also be determined from the Ti\,{\sc ii} lines as they are numerous and
of suitable strength. For a given model, we compute the dispersion in the Fe (or Ti) abundances over
a range in the $\xi$ from 2 to 10 km s$^{-1}$. In Figure~\ref{f_micro}, the dispersion $\sigma$
for Fe and Ti lines is displayed. The dispersion $\sigma$ values for the Fe\,{\sc i} and
Ti\,{\sc ii} lines are computed for two different effective temperature values: $T_{\rm eff}$ =
5750 K (e.g. green curve for Fe) and 6500 K (e.g. blue curve for Ti). From the dispersion
$\sigma$ vs microturbulence velocity $\xi$ plots for the Fe lines, the microturbulence velocity
is found to be in the range 4.7$\leq$$\xi$$\leq$6.0 km s$^{-1}$. A minimum value of $\sigma$ for
the Ti lines is reached at $\xi$ $\approx$ 5.0 km s$^{-1}$. We adopt 4.7$\pm$0.5 km s$^{-1}$.  The
solutions for $\xi$ are not particularly dependent on the chosen $T_{\rm eff}$ of the model.
\vskip 0.2 cm

\noindent Imposition of ionization equilibrium for an element represented by lines from neutral atoms and singley-charged
ions provides a locus in the temperature-gravity plane running from low $T_{\rm eff}$ and low $g$ to high $T_{\rm
eff}$ and high $g$. Such loci are shown in Figure~\ref{f_teff_logg}  for Mg, Si, Ca, Cr, and Fe; silicon presents a problem (see
below). These loci with the (vertical) loci provided by the $T_{\rm eff}$ from excitation of Fe\,{\sc i} lines and the
Paschen line profiles (see below) give $T_{\rm eff}=$6500 K and $\log g=$+0.5 cgs when greatest weight is given to the Fe
ionization equilibrium on account of the greater number of Fe lines.

\subsection{The Balmer and Paschen lines}

\noindent  In principle, Balmer line profiles offer an alternative method of estimating
atmospheric parameters. For warm supergiants like IRAS 18095+2704, the Balmer lines
are sensitive to $T_{\rm eff}$  and insensitive to $\log g$ so that they provide
an isothermal in  Figure~\ref{f_teff_logg}. H$\alpha$ is unsuited to this purpose because
it shows strong emission distorting the expected deep and broad photospheric
absorption profile. Both H$\beta$ and H$\gamma$ appear to have the
anticipated profiles. Predicted profiles are computed with {\sc SYNTHE} as {\sc MOOG} does not compute synthetic
profiles for the hydrogen lines.
Figure~\ref{f_h_beta} shows observed and
predicted profiles from synthetic spectrum calculations for $T_{\rm eff}$
from 5500 K to 6500 K. It is apparent that for both lines the best-fitting
predicted profile corresponds to an effective temperature of around 5750 K; this
estimate is almost independent of the adopted surface gravity. This temperature is significantly
cooler than that suggested by the intersection of the Fe
excitation temperature and the various ionization equilibria (Figure~\ref{f_teff_logg} ).
\vskip 0.2 cm

\noindent Emission in the Balmer lines may be responsible in part for a systematically
lower estimate of the effective temperature. The H$\alpha$ profile
is very strongly affected by emission at the time of our
observation. In order to provide a more complete assessment of the emission, we
use the H$\alpha$ profile illustrated by Khochkova (1995).
The observed profile is much shallower than a profile
synthesized according to the adopted atmospheric parameters (i.e., $T_{\rm eff} = 6500$K).
Obviously, emission is present across the entire profile  and sharp
emission just to the red of center hints at presence of a P Cygni profile.
An emission profile expressed as the difference between the adopted observed
profile and that computed for 6500 K has the shape necessary to correct the observed H$\beta$ 
and H$\gamma$ profiles to the predicted profiles also for 6500 K.
\vskip 0.2 cm

\noindent The outstanding issue is the question of the strength of the emission at H$\beta$ and H$\gamma$.
Standard Case A (where HI in the ionized gas is optically thin; i.e., of low density that Lyman photons can escape) and
Case B (where HI in the ionized gas is optically thick to Lyman line photons, i.e., Lyman line photons in
consequence get absorbed) decrements of the recombination theory for ionized gas suggest that flux ratios might be H$\alpha$/H$\beta$ 
$\sim 3$ and H$\beta$/H$\gamma$ $\sim 2.5$ (Osterbrock \& Ferland 2006). Reduction of the H$\alpha$ emission by
these factors provides an emission component at H$\beta$ and H$\gamma$ that is slightly too weak to reconcile the
observed (corrected for emission) with the profile predicted for 6500 K. Additionally, the fact that H$\gamma$
and H$\beta$ require the same low effective temperature ($T_{\rm eff} \simeq 5750$ K) suggests that superposition of emission on a photospheric profile may not be the explanation. Possibly, the Balmer line profiles are influenced by an upper photosphere (a highly-structured chromosphere with a
developing wind?) that distorts the observed profiles in a way not modellable using Case A or Case B decrements.
\vskip 0.2 cm

\noindent Paschen lines are expected to be less sensitive to structural adjustments of the upper photosphere and to
emission. Figure~\ref{f_paschen} shows observed and synthesized profiles for a selection of Paschen lines recorded on
the McDonald spectrum. Profiles for the model with $T_{\rm eff}$ = 6500 K, $\log g = +0.5$ offer a fine fit to the
observed profiles. The latter are very poorly fit with the $T_{\rm eff} = 5750$K model suggested by the Balmer
H$\beta$ and H$\gamma$ profiles. Thus, the Paschen profiles confirm the effective temperature from the excitation of
the Fe\,{\sc i} lines. The new isothermal provided by the hydrogen lines of the Paschen series at 8862 \AA\  (P$_{\rm
11}$) and 8598 \AA\  (P$_{\rm 14}$) coincides with the
locus from the excitation of Fe\,{\sc i} lines in Figure~\ref{f_teff_logg}.

 \begin{table*}
     \caption[]{Fe I and Fe II lines used in the analysis and corresponding abundances for a model of $T_{\rm eff}=6500$ K, $\log$ g = 0.5. RMT is the multiplet number given by
     the Revised Multiplet Table (Moore 1945).}
        \label{}
    $$
        \begin{array}{@{}ccr@{}c@{}crr||ccr@{}c@{}crr@{}}
           \hline
           \hline
 $Species$&\lambda &  $EW$\,\, &\log\epsilon(Fe)\,\,& $LEP$   &\log(gf)   & RMT&$Species$&\lambda &  $EW$ \,\,&\log\epsilon(Fe)\,\,& $LEP$   &\log(gf)&RMT\\
        & ($\AA$)& (m$\AA$)& (dex)          & ($eV$)  &           &  &	& ($\AA$)& (m$\AA$)& ($dex$)	       & ($eV$)  &  &\\
           \hline
Fe $\,{\sc i}$ &4447.717 &135 & 6.74 &2.22 &-1.342 &68     &Fe $\,{\sc i}$  &5429.697 &220&6.83&0.96&-1.879&15 \\  
Fe $\,{\sc i}$ &4459.118 &153 & 6.78 &2.18 &-1.279 &68     &Fe $\,{\sc i}$  &5434.524 &141&6.42&1.01&-2.122&15 \\
Fe $\,{\sc i}$ &4476.019 &124 & 6.66 &2.85 &-0.819 &350    &Fe $\,{\sc i}$  &5446.917 &206&6.74&0.99&-1.914&15 \\
Fe $\,{\sc i}$ &4531.148 &129 & 6.84 &1.49 &-2.155 &39     &Fe $\,{\sc i}$  &5569.618 & 79&6.43&3.42&-0.486&686\\
Fe $\,{\sc i}$ &4871.318 &193 & 6.81 &2.87 &-0.362 &318    &Fe $\,{\sc i}$  &5572.842 &113&6.47&3.40&-0.275&686\\
Fe $\,{\sc i}$ &4872.138 &139 & 6.54 &2.88 &-0.567 &318    &Fe $\,{\sc i}$  &5615.644 &167&6.54&3.33& 0.050&686\\
Fe $\,{\sc i}$ &4903.308 &85  & 6.46 &2.88 &-0.926 &318	   &Fe $\,{\sc i}$  &5624.542 & 61&6.53&3.42&-0.755&686\\
Fe $\,{\sc i}$ &4918.993 &168 & 6.52 &2.87 &-0.342 &318	   &Fe $\,{\sc i}$  &6230.710 & 68&6.36&2.56&-1.281&207\\
Fe $\,{\sc i}$ &5014.941 & 60 & 6.53 &3.94 &-0.303 &965	   &Fe $\,{\sc i}$  &6393.601 & 56&6.43&2.43&-1.580&168\\
Fe $\,{\sc i}$ &5074.748 & 75 & 6.84 &4.22 &-0.230 &1094   &Fe $\,{\sc i}$  &6411.649 & 53&6.61&3.65&-0.718&816\\
Fe $\,{\sc i}$ &5090.774 & 38 & 6.69 &4.26 &-0.440 &1090   &Fe $\,{\sc ii}$ &4472.929 &154&6.74&2.84&-3.530&37 \\
Fe $\,{\sc i}$ &5216.274 & 98 & 6.66 &1.61 &-2.150 &36     &Fe $\,{\sc ii}$ &4893.820 & 58&6.64&2.83&-4.270&36 \\
Fe $\,{\sc i}$ &5232.940 &224 & 6.86 &2.94 &-0.057 &383    &Fe $\,{\sc ii}$ &5120.352 & 63&6.63&2.83&-4.214&35 \\
Fe $\,{\sc i}$ &5302.302 & 72 & 6.49 &3.28 &-0.720 &553    &Fe $\,{\sc ii}$ &5132.669 & 70&6.55&2.81&-4.090&35 \\
Fe $\,{\sc i}$ &5324.179 &148 & 6.42 &3.21 &-0.103 &553    &Fe $\,{\sc ii}$ &5254.929 &143&6.64&3.23&-3.227&49 \\
Fe $\,{\sc i}$ &5364.871 & 77 & 6.59 &4.45 & 0.228 &1146   &Fe $\,{\sc ii}$ &5256.938 & 71&6.79&2.89&-4.250&41 \\
Fe $\,{\sc i}$ &5367.467 & 93 & 6.49 &4.42 & 0.443 &1146   &Fe $\,{\sc ii}$ &5325.553 &128&6.50&3.22&-3.220&49 \\
Fe $\,{\sc i}$ &5369.962 &105 & 6.45 &4.37 & 0.536 &1146   &Fe $\,{\sc ii}$ &5432.967 & 93&6.68&3.27&-3.629&55 \\
Fe $\,{\sc i}$ &5383.369 &113 & 6.35 &4.31 & 0.645 &1146   &Fe $\,{\sc ii}$ &6147.741 &104&6.40&3.89&-2.721&74 \\
Fe $\,{\sc i}$ &5393.168 & 67 & 6.40 &3.24 &-0.715 &553    &Fe $\,{\sc ii}$ &6149.258 &106&6.54&3.89&-2.840&74 \\
Fe $\,{\sc i}$ &5397.128 &192 & 6.64 &0.92 &-1.993 &15     &Fe $\,{\sc ii}$ &6416.920 & 93&6.47&3.89&-2.880&74 \\
\hline				
\hline
        \end{array}
    $$
\end{table*}

  \begin{table*}
     \caption[]{Species used in the analysis and corresponding abundances for a model of $T_{\rm eff}=6500$ K, $\log$ g = 0.5. RMT is the multiplet number given by
     the Revised Multiplet Table (Moore 1945).}
        \label{}
    $$
        \begin{array}{@{}lcr@{}c@{}rrr||lcr@{}c@{}rrr}
           \hline
           \hline
 $Species$&\lambda &  $EW$\,\, &\log\epsilon(X)\,\,& $LEP$   &\log(gf)   &RMT& $Species$&\lambda &  $EW$ \,\,&\log\epsilon(X)\,\,& $LEP$   &\log(gf)&RMT\\
          & ($\AA$)& (m$\AA$)  & (dex)              & ($eV$)  &           & 	&   & ($\AA$)& (m$\AA$)	&   ($dex$)                 &($eV$)  & &\\
           \hline
                                                                
C $\,{\sc i}$  &4766.620 & 40 & 8.28 & 7.48 &-2.62  & 6    & Ca $\,{\sc ii}$&5019.971 &SS   & 5.40  & 7.52 &-0.28 &15  \\
C $\,{\sc i}$  &4770.000 & 27 & 7.90 & 7.48 &-2.44  & 6    & Ca $\,{\sc ii}$&5285.266 &14   &5.43   & 7.51 &-1.18 &14  \\    
C $\,{\sc i}$  &4771.690 &108 & 8.15 & 7.49 &-1.87  & 6    & Ca $\,{\sc ii}$&5307.224 &SS   &5.50   & 7.52 &-0.90 &14  \\     
C $\,{\sc i}$  &4775.870 & 51 & 8.10 & 7.49 &-2.30  & 6    & Ca $\,{\sc ii}$&5339.190 & 12  &5.25   & 8.44 &-0.33 &20  \\ 
C $\,{\sc i}$  &4932.049 & 64 & 7.75 & 7.68 &-1.66  &13	   & Sc $\,{\sc ii}$&5031.019 &169  &2.06   & 1.36 &-0.40 &23  \\ 
C $\,{\sc i}$  &5380.340 & SS & 7.85 & 7.68 &-1.62  &11    & Sc $\,{\sc ii}$&5239.823 & 78  &1.82   & 1.46 &-0.77 &26  \\
C $\,{\sc i}$  &7111.470 & SS & 7.90 & 8.64 &-1.09  &26    & Sc $\,{\sc ii}$&5318.350 & 19  &2.23   & 1.36 &-2.01 &22  \\
C $\,{\sc i}$  &7113.180 & SS & 7.80 & 8.65 &-0.77  &26    & Sc $\,{\sc ii}$&5526.810 &164  &1.95   & 1.77 & 0.02 &31  \\
C $\,{\sc i}$  &7115.170 & SS & 7.77 & 8.64 &-0.94  &26    & Sc $\,{\sc ii}$&5640.971 & 69  &2.13   & 1.50 &-1.13 &29  \\
C $\,{\sc i}$  &7116.990 & SS & 7.80 & 8.65 &-0.91  &26    & Sc $\,{\sc ii}$&5667.149 & SS  &\leq1.94&1.50 &-1.31 &29  \\
C $\,{\sc i}$  &7119.660 & SS & 7.85 & 8.64 &-1.15  &26    & Sc $\,{\sc ii}$&5669.040 & 64  &2.15   & 1.50 &-1.20 &29  \\
N $\,{\sc i}$  &8216.340 & SS &\leq7.19&10.34& 0.13 &2     & Sc $\,{\sc ii}$&5684.190 & 90  &2.25   & 1.51 &-1.07 &29  \\
N $\,{\sc i}$  &8223.130 & SS &\leq7.19&10.33&-0.27 &2	   & Ti $\,{\sc i}$ &4999.503 & 33  &3.97   & 0.83 & 0.25 &38  \\
$[O\,{\sc i}]$ &5577.340 & SS & 8.52 & 1.97 &-8.23  &3	   & Ti $\,{\sc i}$ &5036.468 & SS  &\leq4.13& 1.44 & 0.13 &110\\
$[O\,{\sc i}]$ &6300.304 & 18 & 8.57 & 0.00 &-9.78  &1	   & Ti $\,{\sc i}$ &5192.969 & SS  &\leq4.15& 0.02 &-1.01 &4  \\ 
$[O\,{\sc i}]$ &6363.780 & SS & 8.42 & 0.02 &-10.26 &1     & Ti $\,{\sc ii}$&4493.513 & 72  &4.10   & 1.08 &-3.02 &18  \\
O $\,{\sc i}$  &6155.970 & SS & 8.60 &10.74 &-0.66  & 10   & Ti $\,{\sc ii}$&4544.009 & 102 &4.04   & 1.24 &-2.58 &60  \\
O $\,{\sc i}$  &6156.760 & SS & 8.52 &10.74 &-0.44  & 10   & Ti $\,{\sc ii}$&4568.312 &  78 &4.19   & 1.22 &-2.94 &60  \\
O $\,{\sc i}$  &6158.170 & SS & 8.67 &10.74 &-0.30  & 10   & Ti $\,{\sc ii}$&4708.663 & 153 &4.16   & 1.24 &-2.34 &48  \\
Na $\,{\sc i}$ &4978.541 & 28 & 6.15 & 2.10 &-1.22  &9	   & Ti $\,{\sc ii}$&4798.535 & 120 &4.11   & 1.08 &-2.68 &17  \\
Na $\,{\sc i}$ &5682.633 & SS & 6.02 & 2.10 &-0.71  & 6    & Ti $\,{\sc ii}$&4865.620 & 128 &4.30   & 1.11 &-2.79 &29  \\
Na $\,{\sc i}$ &5688.205 & SS & 5.87 & 2.10 &-0.45  & 6    & Ti $\,{\sc ii}$&4874.014 & 114 &3.96   & 3.09 &-0.80 &114 \\
Na $\,{\sc i}$ &8183.255 & SS & 6.27 & 2.10 & 0.24  & 4    & Ti $\,{\sc ii}$&4911.193 & 143 &4.03   & 3.12 &-0.61 &114 \\
Na $\,{\sc i}$ &8194.790 & SS & 6.27 & 2.10 &-0.46  & 4    & Ti $\,{\sc ii}$&5072.281 & 113 &4.24   & 3.12 &-1.06 &113 \\
Mg $\,{\sc i}$ &4057.505 & SS & 7.22 & 4.35 &-0.90  &16    & Ti $\,{\sc ii}$&5185.900 & 173 &4.03   & 1.89 &-1.49 &86  \\
Mg $\,{\sc i}$ &4702.991 &205 & 7.10 & 4.35 &-0.44  &11    & Ti $\,{\sc ii}$&5381.015 & SS  &3.73   & 1.57 &-1.92 &69  \\
Mg $\,{\sc i}$ &5528.410 &217 & 7.26 & 4.35 &-0.50  &9     & Ti $\,{\sc ii}$&5418.751 & 92  &3.63   & 1.58 &-2.00 &69  \\
Mg $\,{\sc i}$ &5711.090 & 47 & 6.98 & 4.34 &-1.72  &8     & V  $\,{\sc ii}$&4036.779 &  SS &3.14   & 1.48 &-1.53 &9   \\
Mg $\,{\sc ii}$&4481.126 & SS & 7.17 & 8.86 & 0.75  &4     & V  $\,{\sc ii}$&4039.574 &  SS &\leq3.14& 1.82&-2.11 &32  \\
Mg $\,{\sc ii}$&4481.150 & SS & 7.17 & 8.86 &-0.55  &4     & V  $\,{\sc ii}$&4564.578 &  SS &3.34   &2.27 &-1.39 &56  \\
Mg $\,{\sc ii}$&4481.325 & SS & 7.17 & 8.86 & 0.59  &4     & Cr $\,{\sc i}$ &4545.945 &  25 &4.80   & 0.94 &-1.38 &10  \\
Al $\,{\sc i}$ &3944.020 & SS & 4.91 & 0.00 &-0.64  &1	   & Cr $\,{\sc i}$ &4646.144 &  51 &4.57   & 1.03 &-0.71 &21  \\
Al $\,{\sc i}$ &3961.523 & SS & 4.81 & 0.01 &-0.34  &1     & Cr $\,{\sc i}$ &4652.158 &  29 &4.57   & 1.00 &-1.03 &21  \\
Si $\,{\sc i}$ &5665.550 & 12 & 7.06 & 4.92 &-1.94  &10    & Cr $\,{\sc i}$ &5204.518 & 142 &4.71   & 0.94 &-0.21 &7   \\
Si $\,{\sc i}$ &5701.108 & 19 & 7.29 & 4.93 &-1.95  &10    & Cr $\,{\sc i}$ &5208.420 & 197 &4.82   & 0.94 & 0.16 &7   \\
Si $\,{\sc i}$ &5708.437 & 46 & 7.17 & 4.95 &-1.37  &10    & Cr $\,{\sc i}$ &5296.686 &  SS &4.81   & 0.98 &-1.41 &18  \\
Si $\,{\sc i}$ &5772.150 & 20 & 7.14 & 5.08 &-1.65  &17	   & Cr $\,{\sc i}$ &5297.360 &  SS &4.81   & 2.90 & 0.17 &94  \\
Si $\,{\sc i}$ &5793.070 & 29 & 7.50 & 4.93 &-1.96  &9	   & Cr $\,{\sc i}$ &5298.269 &  SS &4.71   & 0.98 &-1.16 &18  \\
Si $\,{\sc i}$ &6145.020 & SS &\leq6.99& 5.62&-0.84 &29    & Cr $\,{\sc i}$ &5348.312 &  18 &4.56   & 1.00 &-1.29 &18  \\
Si $\,{\sc i}$ &6155.140 & 66 & 6.90 & 5.62 &-0.30  &29    & Cr $\,{\sc i}$ &5409.790 &  68 &4.71   & 1.03 &-0.72 &18  \\
Si $\,{\sc ii}$&5055.980 & SS & 6.70 &10.07 &0.52   &5	   & Cr $\,{\sc ii}$&4812.350 &  68 &4.42   & 3.86 &-1.80 &30  \\
Si $\,{\sc ii}$&6347.091 &285 & 8.03 & 8.12 & 0.15  &2     & Cr $\,{\sc ii}$&4884.607 &  57 &4.58   & 3.86 &-2.08 &30  \\
Si $\,{\sc ii}$&6371.359 &252 & 7.91 & 8.12 &-0.08  &2     & Cr $\,{\sc ii}$&5246.768 &  35 &4.56   & 3.71 &-2.46 &23  \\
S  $\,{\sc i}$ &4694.113 & 43 & 6.98 & 6.53 &-1.71  &2     & Cr $\,{\sc ii}$&5305.850 &  57 &4.55   & 3.83 &-2.08 &24  \\
S  $\,{\sc i}$ &4695.443 & 39 & 7.08 & 6.53 &-1.87  &2     & Cr $\,{\sc ii}$&5308.440 &  53 &4.44   & 4.07 &-1.81 &43  \\
S  $\,{\sc i}$ &4696.252 & 16 & 6.88 & 6.53 &-2.10  &2     & Cr $\,{\sc ii}$&5310.687 &  28 &4.57   & 4.07 &-2.27 &43  \\
S  $\,{\sc i}$ &6052.583 & SS & 7.00 & 7.86 &-1.26  &10    & Cr $\,{\sc ii}$&5313.590 &  98 &4.69   & 4.07 &-1.65 &43  \\
S  $\,{\sc i}$ &6743.580 & 25 & 7.20 & 7.87 &-1.07  &8     & Cr $\,{\sc ii}$&5478.365 &  52 &4.62   & 4.17 &-1.91 &50  \\
S  $\,{\sc i}$ &6748.840 & SS & 6.90 & 7.87 &-0.64  &8     & Cr $\,{\sc ii}$&5510.702 &  27 &4.51   & 3.82 &-2.45 &23  \\
S  $\,{\sc i}$ &6757.170 & SS & 6.83 & 7.87 &-0.35  &8     & Mn $\,{\sc i}$ &4041.361 &  SS &4.48   & 2.11 & 0.29 &5   \\
Ca $\,{\sc i}$ &4425.441 & 95 & 5.36 & 1.88 &-0.36  &4     & Mn $\,{\sc i}$ &4754.040 &  43 &4.36   & 2.27 &-0.09 &16  \\
Ca $\,{\sc i}$ &4434.957 & SS & 5.50 & 1.88 &-0.01  &4     & Mn $\,{\sc i}$ &4783.420 &  64 &4.47   & 2.29 & 0.04 &16  \\
Ca $\,{\sc i}$ &4435.679 & SS & 5.50 & 1.88 &-0.52  &4     & Co $\,{\sc i}$ &3995.306 &  SS &4.26   & 0.92 &-0.22 &31  \\
Ca $\,{\sc i}$ &4578.550 & 19 & 5.21 & 2.52 &-0.56  &23    & Co $\,{\sc i}$ &4121.318 &  SS &\leq4.06	& 0.92 &-0.32 &28  \\
Ca $\,{\sc i}$ &5581.965 & SS & 5.60 & 2.52 &-0.71  &21    & Ni $\,{\sc i}$ &4714.408 &  78 &5.34   & 3.38 & 0.23 &98  \\
Ca $\,{\sc i}$ &5588.749 & 98 & 5.35 & 2.53 & 0.21  &21    & Ni $\,{\sc i}$ &4829.016 &  26 &5.43   & 3.54 &-0.33 &131 \\
Ca $\,{\sc i}$ &5590.114 & SS & 5.60 & 2.52 &-0.71  &21    & Ni $\,{\sc i}$ &4937.341 &  19 &5.40   & 3.61 &-0.40 &114 \\
Ca $\,{\sc i}$ &5594.462 & 80 & 5.45 & 2.52 &-0.05  &21    & Ni $\,{\sc i}$ &4980.166 &  42 &5.50   & 3.61 &-0.11 &112 \\
Ca $\,{\sc i}$ &5601.277 & 33 & 5.59 & 2.53 &-0.69  &21    & Ni $\,{\sc i}$ &5035.357 &  SS &5.39   & 3.64 & 0.29 &143 \\
Ca $\,{\sc i}$ &5857.454 & 73 & 5.47 & 2.93 & 0.23  &47    & Ni $\,{\sc i}$ &5080.523 &  SS &5.39   & 3.66 & 0.13 &143 \\
Ca $\,{\sc i}$ &6122.219 &131 & 5.57 & 1.89 &-0.32  &3     & Ni $\,{\sc i}$ &5081.107 &  SS &5.54   & 3.85 & 0.30 &194 \\
Ca $\,{\sc i}$ &6162.172 &161 & 5.60 & 1.90 &-0.09  &3     & Ni $\,{\sc i}$ &5082.354 &  SS &5.69   & 3.66 &-0.54 &130 \\
Ca $\,{\sc i}$ &6462.566 &122 & 5.44 & 2.52 & 0.31  &18    & Ni $\,{\sc i}$ &5084.081 &  SS &5.69   & 3.68 & 0.03 &162 \\
\hline
\hline
        \end{array}
    $$
\end{table*}

  \begin{table*}
     \caption[]{(Continued.)}
        \label{}
    $$
        \begin{array}{@{}lcr@{}c@{}rrr||lcr@{}c@{}rrr}
           \hline
           \hline
 $Species$&\lambda &  $EW$\,\, &\log\epsilon(X)\,\,& $LEP$   &\log(gf)   &RMT& $Species$&\lambda &  $EW$ \,\,&\log\epsilon(X)\,\,& $LEP$   &\log(gf)&RMT\\
          & ($\AA$)& (m$\AA$)  & (dex)              & ($eV$)  &           & 	&   & ($\AA$)& (m$\AA$)	&   ($dex$)                 &($eV$)  & &\\
           \hline
                                                          
Ni $\,{\sc i}$ &5099.927 &  SS &5.44   & 3.68 &-0.10 &161 &Y  $\,{\sc ii}$&5119.111 &  SS &\leq0.81& 0.99&-1.36 &20  \\
Ni $\,{\sc i}$ &5155.762 &  SS &5.44   & 3.90 &-0.09 &210 &Y  $\,{\sc ii}$&5123.210 &  SS &0.81   &0.99  &-0.83 &21  \\
Cu $\,{\sc i}$ &5105.537 &  SS &\leq3.35& 1.39&-1.52 &2   &Y  $\,{\sc ii}$&5200.413 &  SS &\leq0.68& 0.99&-0.57 &20  \\
Zn $\,{\sc i}$ &4680.138 &  19 &3.91   & 4.00 &-0.86 &2   &Zr $\,{\sc ii}$&4050.329 &  SS &\leq1.24&0.71 &-1.06 &43  \\
Zn $\,{\sc i}$ &4722.153 &  44 &3.88   & 4.03 &-0.39 &2   &Zr $\,{\sc ii}$&4208.980 &  SS &1.14   & 0.71 &-0.51 &41  \\
Zn $\,{\sc i}$ &4810.534 &  53 &3.80   & 4.08 &-0.17 &2   &Zr $\,{\sc ii}$&4211.880 &  44 &1.35   & 0.53 &-1.04 &15  \\
Sr $\,{\sc ii}$&4077.714 & SS  & 2.54  & 0.00 & 0.14 & 1  &Zr $\,{\sc ii}$&4496.960 &  54 &1.44  & 0.71&-0.89&40     \\
Sr $\,{\sc ii}$&4215.524 & SS  & 2.49  & 0.00 &-0.18 & 1  &Ba $\,{\sc ii}$&5853.675 &  79 &1.05  & 0.60 &-1.01 &2    \\
Y  $\,{\sc ii}$&4854.867 &  SS &0.88   & 0.99 &-0.38 &22  &La $\,{\sc ii}$&5114.560 &  SS &\leq0.31& 0.24&-1.03 &36  \\
Y  $\,{\sc ii}$&4883.690 &  83 &0.59   & 1.08 & 0.07 &22  &Nd $\,{\sc ii}$&4303.580 &  SS &\leq0.44&0.00&0.08 &10    \\
Y  $\,{\sc ii}$&5087.420 &  SS &0.58   &1.08  &-0.17 &20  &Eu $\,{\sc ii}$&4129.720 &  SS &-0.05  & 0.00 &0.22  &1   \\
\hline
\hline
        \end{array}
    $$
\end{table*}

 \begin{table*}
    \caption[]{Abundances of the observed species for IRAS 18095+2704 are presented
    for a model atmospheres of $T_{\rm eff} = 6700$ K, $\log$ g = 0.3 (KT), $\xi$ = 6.0 and $T_{\rm eff}=6500$ K, $\log$ g = 0.5, $\xi$
    = 4.7 (this work). The solar abundances from
    Asplund et al. (2009) is used to
    convert Klochkova's abundance of element X to [X/Fe] for comparison purposes.}
       \label{}
   $$
       \begin{array}{l||cc||@{}c@{}c@{}lc||c}
          \hline
          \hline
%               & & (T_{\rm eff}, \log\,g) &     \\
 $Species$     &K95 &K95 &   & \,\,\,\,This\,work &  & &\log\epsilon_{\odot}   \\
\cline{2-7}
%\cline{3-3}
               &\log\epsilon(X) & $[X/Fe]$ & \,\,\log\epsilon(X) & $[X/Fe]$& \,\,\,\Delta	&N  &  \\
          \hline
          \hline
 C$\,{\sc i}$   &8.27&+0.50&7.92&+0.40 & -0.35 & 11&8.43  \\
 N$\,{\sc i}$   &7.66&+0.47&\leq7.19&\leq+0.27 & -0.47\leq& 2 &7.83  \\
 O$\,{\sc i}$   &8.74&+0.65&8.55&+0.77 & -0.19 &6 &8.69  \\
 Na$\,{\sc i}$  &6.02&+0.47&6.12&+0.79 &+0.10  &5 &6.24  \\
 Mg$\,{\sc i}$  &7.42&+0.62&7.14&+0.45 &-0.28  &4 &7.60  \\
 Mg$\,{\sc ii}$ &7.52&+0.72&7.17&+0.48 &-0.35  &1 &7.60  \\ 
 Al$\,{\sc i}$  &5.81&+0.12&4.86&-0.68 &-0.95  &2 &6.45  \\
 Si$\,{\sc i}$  &7.48&+0.71&7.18&+0.58 &-0.30  &6 &7.51  \\
 Si$\,{\sc ii}$ &... & ... &6.70&+0.10 &...    &1 &7.51  \\
 S$\,{\sc i}$   &6.96&+0.53&6.98&+0.77 &+0.02  &7 &7.12  \\
 Ca$\,{\sc i}$  &5.84&+0.26&5.48&+0.05 &-0.36  &13&6.34  \\
 Ca$\,{\sc ii}$ &... & ... &5.40&-0.03 &...    &4 &6.34  \\
 Sc$\,{\sc ii}$ &2.25&-0.30&2.08&-0.16 &-0.17  &7 &3.15  \\
 Ti$\,{\sc i}$  &5.68&+1.44&3.97&-0.07 &-1.71  &1 &4.95  \\
 Ti$\,{\sc ii}$ &4.05&-0.19&4.04&+0.00 &-0.01  &12&4.95  \\
 V$\,{\sc ii} $ &3.45&+0.23&3.24&+0.22 &-0.21  &2 &3.93  \\
 Cr$\,{\sc i}$  &6.02&+1.13&4.71&-0.02 &-1.31  &10&5.64  \\
 Cr$\,{\sc ii}$ &4.98&+0.09&4.54&-0.19 &-0.44  &9 &5.64  \\
 Mn$\,{\sc i}$  &5.12&+0.51&4.44&-0.08 &-0.68  &3 &5.43  \\
 Fe$\,{\sc i}$  &6.71&-0.01&6.59&+0.00 &-0.12  &31&7.50  \\
 Fe$\,{\sc ii}$ &6.73&+0.01&6.59&+0.00 &-0.14  &11&7.50  \\
 Co$\,{\sc i} $ &5.74&+1.60&4.26&+0.18 &-1.48  &1 &4.99  \\
 Ni$\,{\sc i} $ &6.13&+0.66&5.48&+0.17 &-0.65  &11&6.22  \\
 Cu$\,{\sc i } $&4.06&+0.63&\leq3.35&\leq+0.07 &-0.71\leq &1 &4.19  \\
 Zn$\,{\sc i} $ &4.60&+0.78&3.86&+0.21 &-0.74  &3 &4.56  \\
 Sr$\,{\sc ii}$ &... &...  &2.52&+0.56 &...    &2 &2.87  \\
 Y $\,{\sc ii}$ &1.42&-0.04&0.72&-0.58 &-0.70  &4 &2.21  \\
 Zr$\,{\sc ii} $&... & ... &1.31&-0.36 &...    &3 &2.58  \\
 Ba$\,{\sc ii}$ &1.13&-0.22&1.05&-0.22 &-0.08  &1 &2.18  \\
 La$\,{\sc ii}$ &0.83&+0.39&\leq0.31&\leq+0.12 &-0.52\leq & 1 &1.10  \\
 Nd$\,{\sc ii}$ &1.56&+0.84&\leq0.44&\leq-0.07 & -1.12\leq& 1 &1.42  \\
 Eu$\,{\sc ii}$ &0.96&+1.23&-0.05&+0.34&-1.01  & 1 &0.52  \\
\hline
\hline
       \end{array}
   $$
\begin{list}{}{}
\item  \hskip 4.0 cm Notes. K95: Klochkova (1995).
\item  \hskip 4.8 cm $\Delta$ =  $\log\epsilon$(X)$_{This\,work}$- $\log\epsilon$(X)$_{K95}$
\item  \hskip 4.8 cm N is the number of the lines employed in our abundances determination.
\item  \hskip 4.8 cm The solar abundances ($\log\epsilon_{\odot}$) are from Asplund et al. (2009). 
\end{list}
 \end{table*}

\section{ABUNDANCE ANALYSIS -- Elements and Lines}

\noindent For the lines of neutral and/or singly-ionized atoms, we conducted a systematic search by using lower
excitation potential and $gf$-value as the guides. The {\sc Revised Multiplet Table (RMT)} (Moore 1945) is used
as an initial guide in this basic step. Lines chosen by this search are listed in Table 3. When a reference to
solar abundances is necessary in order to convert our abundance of element X to either of the quantities [X/H]
or [X/Fe], Asplund et al. (2009) is preferred.
\vskip 0.2 cm

\noindent Comments on individual elements with notes on the adopted $gf$-values follow:
\vskip 0.2 cm

\begin{figure}
 \centering
 \includegraphics[width=0.99\columnwidth,height=75mm,angle=0]{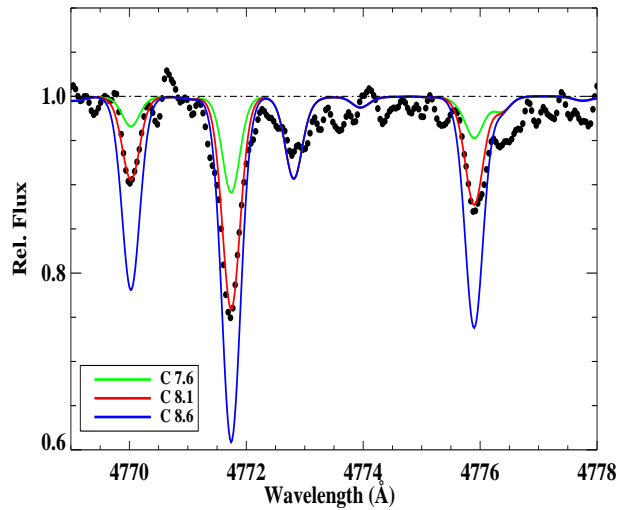}
\caption{The observed spectrum near 4774 \AA\ and synthetic spectra for the three carbon
abundances shown on the figure. }
     \label{CI_4771_synthesis}
      \end{figure}

\noindent {\bf C:} The $gf$-values are taken from Wiese, F\"{u}hr \& Deter (1996).
Ten lines give a mean abundance
$\log\epsilon$(C)=7.92$\pm$0.17. Observed and synthetic spectra for three
different carbon abundances are shown in Figure~\ref{CI_4771_synthesis} for a
region providing three of the ten chosen lines.
\vskip 0.2 cm

\begin{figure}
 \centering
 \includegraphics[width=0.99\columnwidth,height=75mm,angle=0]{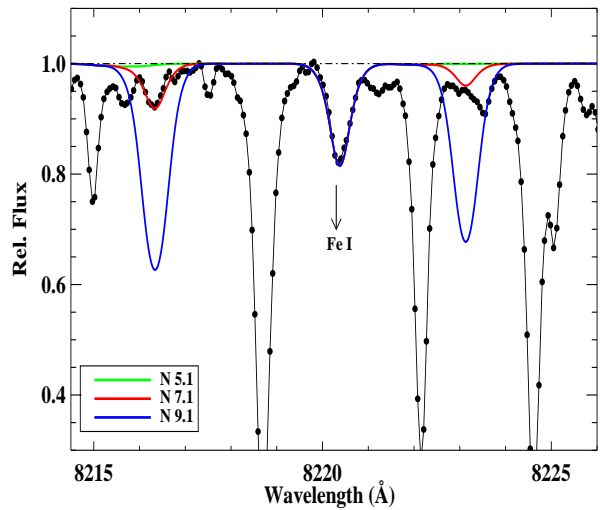}
\caption{The observed spectrum near 8216\AA\ and synthetic spectra  for the three
nitrogen abundances shown on the figure.}
      \label{f_n_8216_synthesis}
       \end{figure}

\noindent {\bf N:}  The $gf$-values are taken from
Wiese, F\"{u}hr \& Deter (1996). Promising N\,{\sc i} lines are in the red outside the
wavelength range covered by the SAO spectrum. The McDonald spectrum covers the
region spanned by several multiplets.  Several potential lines fall in inter-order
gaps.
Figure~\ref{f_n_8216_synthesis} shows a possible detection of one N\,{\sc i} line.
The N abundance is
$\log\epsilon$(N)$\leq 7.1$ but this might properly be considered an upper
limit. 
\vskip 0.2 cm

\begin{figure}
 \centering
 \includegraphics[width=0.99\columnwidth,height=75mm,angle=0]{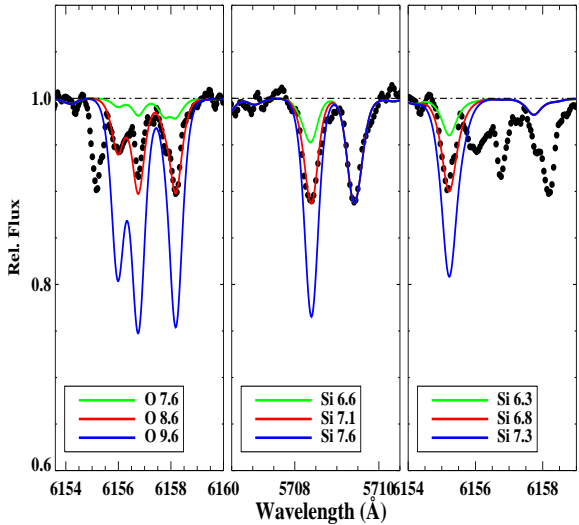}
\caption{The spectrum near 6156\AA, a region containing the Si\,{\sc i} 6155\AA\ and the
three lines of O\,{\sc i} RMT10. Synthetic spectra are shown for the O and Si
abundances indicated on the figure.}
     \label{f_o_synthesis}
      \end{figure}

\noindent {\bf O: } The $gf$-values for the permitted and forbidden O\,{\sc i} lines
are taken from Wiese, F\"{u}hr \& Deter (1996). On the McDonald spectrum, the
forbidden oxygen lines at 5577 \AA, 6300 \AA\ and 6363 \AA\ are detected.
Weak permitted lines of RMT 10 near 6156\AA\ are also analyzed. Figure~\ref{f_o_synthesis} shows the best-fitting
synthetic spectrum for the O\,{\sc i} 6156 \AA\ region.
Forbidden and permitted lines give a very similar
abundance.
The strong O\,{\sc i} triplet at 7774 \AA\ and the 8446 \AA\  feature
give an abundance about 1.5 dex higher abundance, a difference attributed to non-LTE effects.
\vskip 0.2 cm

\noindent {\bf Na:} The $gf$-values
are taken from the {\sc NIST} database.
Five Na\,{\sc i} lines were suitable for abundance analysis from the McDonald
spectrum (Table 3).  RMT 4 and 6  give somewhat different results but we
assign  all lines the same weight.
\vskip 0.2 cm

\begin{figure}
 \centering
 \includegraphics[width=0.99\columnwidth,height=75mm,angle=0]{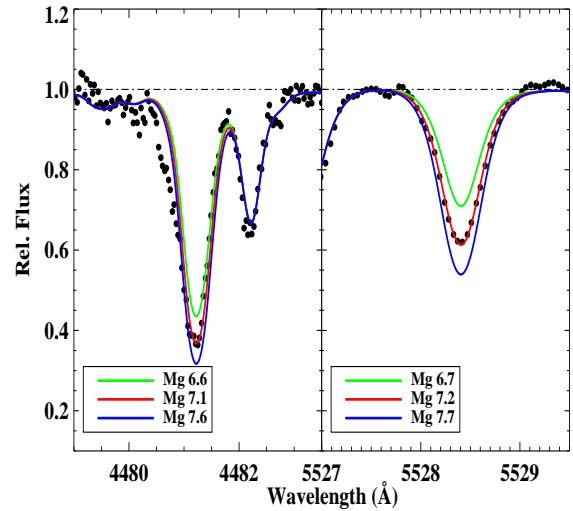}
\caption{The Mg\,{\sc i} 5528 \AA\  and Mg\,{\sc ii} 4481 \AA\ regions showing a comparison of the observed spectrum versus the
synthetic
spectra for three Mg abundances. }
     \label{f_mg_synthesis}
 \end{figure}

\noindent {\bf Mg:} The $gf$-values for Mg\,{\sc i} and Mg\,{\sc ii} lines are taken from the NIST
database.
Four Mg\,{\sc i} lines are listed in Table 3.
Figure~\ref{f_mg_synthesis} shows the best-fitting synthetic spectrum for the
5528 \AA\ Mg\,{\sc i} line. The strongest lines not included in Table 3 are the Mg b triplet at
5167 (blended with Fe\,{\sc i}), 5172
(blended with Fe\,{\sc i}), and 5183 \AA\ have measured equivalent widths of 470
m\AA\ , 413 m\AA\ , and 496 m\AA\ in the McDonald spectrum, respectively; these yield
an (uncertain) abundance of 7.12 dex in good agreement, however, with lines in Table 3.
The strong Mg\,{\sc ii} 4481\AA\ feature (Figure~\ref{f_mg_synthesis}) is well
reproduced by the abundance from the Mg\,{\sc i} lines and shows the asymmetry in
the blue wing attributed to a wind.
\vskip 0.2 cm

\begin{figure}
 \centering
 \includegraphics[width=0.99\columnwidth,height=75mm,angle=0]{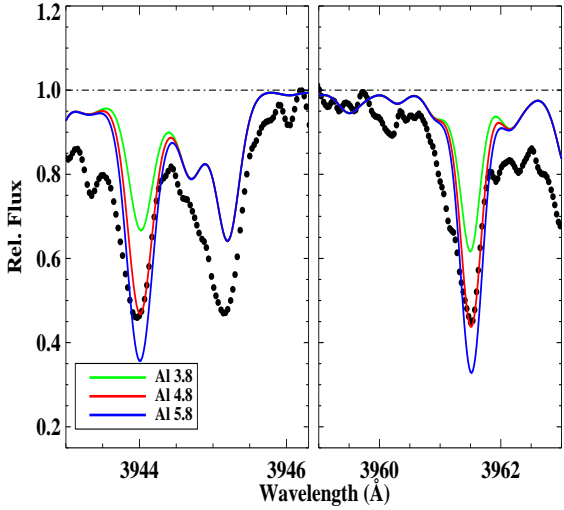}
\caption{Sample region of the McD spectrum including Al\,{\sc i} lines at 3944 \AA\ and 3961 \AA\ .}
      \label{f_al_synth}
 \end{figure}

\noindent {\bf Al:} The resonance lines at 3944 \AA\ and 3961 \AA\ are  detectable.
Their $gf$-values are from the
NIST database.
Comparison of observed and synthetic spectra is presented in Figure~\ref{f_al_synth}.
Excited Al\,{\sc i} lines were searched for but not surprisingly were undetectable.
\vskip 0.2 cm

\noindent {\bf Si:} Selection of Si\,{\sc i} lines
was made starting with the list of lines used by Asplund (2000) for his
solar abundance determination. Asplund's
adopted $gf$-values come from Garz (1973) with the adjustment recommended
by Becker et al. (1980). Asplund et al. (2009)  remark that use of a 3D model
solar atmosphere and non-LTE corrections (Shi et al. 2008) do not change his
2000 estimate for the Si abundance.

\noindent The $gf$-values for Si\,{\sc ii} lines are those recommended by
Kelleher \& Podobedova (2008). Especially prominent in the McDonald spectrum
are the lines at 6347 \AA\ and 6371 \AA. These lines provide an implausibly high
abundance, a value about
1.0 dex higher than that from the Si\,{\sc i} lines and corresponding to
[Si/Fe] $\simeq +1.4$. This abundance is likely an indication that the
lines are not formed in LTE.
A search for weak Si\,{\sc ii} lines yielded the line at
5055.98 \AA\ from  RMT 5 lines which provides the 
abundance $\log\epsilon$(Si) $= 6.7$. This is not only 1.3 dex less
than the value from the 6347 \AA\ and 6371 \AA\ lines but about 0.5 
dex less than the
abundance from Si\,{\sc i } lines.
\vskip 0.2 cm
\noindent {\bf S:} The $gf$-values for S\,{\sc i} lines were taken from
Podobedova et al. (2009).
Seven lines from three multiplets are
easily measurable.
\vskip 0.2 cm

\begin{figure}
 \centering
 \includegraphics[width=0.99\columnwidth,height=75mm,angle=0]{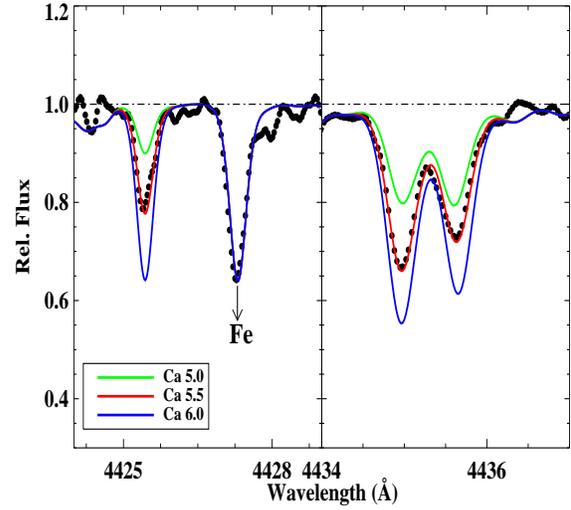}
\caption{The Ca\,{\sc i} 4425 \AA\ and 4435 \AA\ regions showing a comparison of the observed spectrum versus synthetic spectra for three Ca abundances.}
     \label{f_ca_synthesis}
      \end{figure}

\noindent {\bf Ca:} The $gf$-values for the 13 Ca\,{\sc i} lines in Table 3
are taken from the NIST database. New measurements for RMT 3 by Aldenius et al. (2009)
are smaller by only 0.07 dex, a difference that is ignored here. Figure~\ref{f_ca_synthesis} shows the best-fitting
synthetic spectrum for the Ca\,{\sc i} 4425 \AA\ and 4435 \AA\ regions.
\vskip 0.2 cm

\begin{figure}
 \centering
 \includegraphics[width=0.99\columnwidth,height=75mm,angle=0]{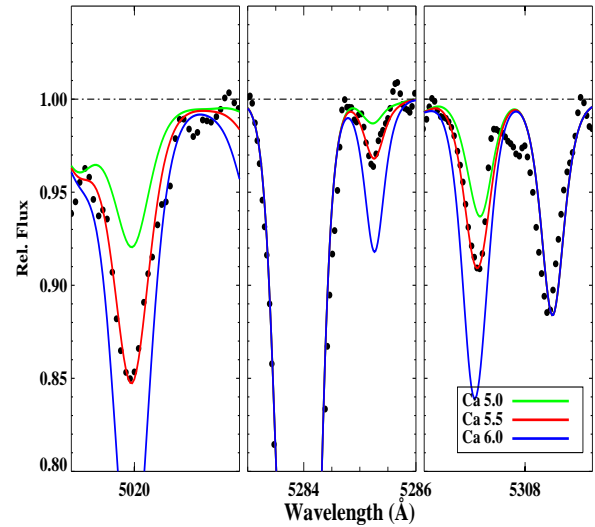}
\caption{The Ca\,{\sc ii} 5019, 5285, and 5307 \AA\ regions showing a comparison of the observed spectrum versus the
synthetic spectra for three Ca abundances.}
     \label{f_ca2_synthesis}
      \end{figure}

\noindent The $gf$-values for three of the four Ca\,{\sc ii} lines in Table 3  are
taken from the NIST database with the entry for the 5339 \AA\ line from the Kurucz database
in the absence of an entry in the former database.
The four lines  give consistent results:
Figure~\ref{f_ca2_synthesis} shows the best-fitting synthetic spectrum for
the Ca\,{\sc ii} lines at 5019, 5285, and 5307 \AA.
\vskip 0.2 cm

\noindent {\bf Sc:} The $gf$-values for the Sc\,{\sc ii} lines in Table 3 are exclusively from
Lawler \& Dakin (1989) who combined radiative lifetime and branching ratio
measurements.
\vskip 0.2 cm
\noindent {\bf Ti:} The $gf$-values for Ti\,{\sc ii} lines are taken from Pickering et al. (2001, 2002). The  4999
\AA\ Ti\,{\sc i} line gives the abundance in Table 3. Neutral titanium abundance is constrained with Ti\,{\sc i} lines
at 5036 \AA\ and 5192 \AA\ for which we set abundance limits in Table 3 using $gf$-values
from the NIST database.
\vskip 0.2 cm

% \begin{figure}
% \centering
% \includegraphics[width=0.99\columnwidth,height=75mm,angle=0]{TiII_5186_synthesis.ps}
%\caption{The Ti\,{\sc ii} 5186 \AA\  and 4568 \AA\ regions showing a comparison of the observed spectrum vs. the best-fitting synthetic spectrum. The macroturbulent velocity used in the synthesis is
%25 km s$^{-1}$. The asymmetry in the 4568 \AA\ line can be clearly seen. The cause for the asymmetry in this low excitation line is,
%probably, due to blending.}
%      \label{f_ti_synthesis}
%       \end{figure}

\noindent {\bf V:} The $gf$-values for V\,{\sc ii} lines are taken from Bi\'{e}mont et al. (1989).
The leading lines in the solar spectrum  expected in the spectrum of {\sc IRAS18095+2704}
are at 4036.77\AA\ and 4564.58\AA\ .
The former is detectable but the latter's absence provides an upper limit to the V 
abundance. 
\vskip 0.2 cm

% \begin{figure}
% \centering
% \includegraphics[width=0.99\columnwidth,height=75mm,angle=0]{VII_4036_4039_synthesis.ps}
%\caption{The V\,{\sc ii} 4036 \AA\ and 4039 \AA\ regions showing a comparison of the observed spectrum vs. the best-fitting synthetic spectrum. The macroturbulent velocity used in the synthesis is
%20 km s$^{-1}$.}
%      \label{f_v_synthesis}
%       \end{figure}

\noindent {\bf Cr:} The $gf$-values for Cr\,{\sc i} lines 
 are taken from the NIST database. Sobeck et al.'s (2007) measurements
are within $\pm$0.02 dex of NIST values for most lines in Table 3 and always within
$\pm$0.10 dex.

\noindent A majority of the selected Cr\,{\sc ii} lines has $gf$-values in the NIST database and for
the missing minority we take semi-empirical values from the Kurucz linelist. Nilsson et al.
(2006) report new measurements of Cr\,{\sc ii} $gf$-values from a combination of
radiative lifetimes and branching fractions. The majority of the chosen stellar lines in the paper are
longward of 4850 \AA\ which is the long wavelength limit for the lines measured by Nilsson
et al. For the seven lines in Nilsson et al.'s list between 4000\AA\ and 4850\AA,
the mean difference between the NIST and their entries for $\log gf$ is just $-0.06$.
Therefore, we make no adjustment to the NIST (and Kurucz) entries.
\vskip 0.2 cm

% \begin{figure}
% \centering
% \includegraphics[width=0.99\columnwidth,height=75mm,angle=0]{CrI_5204_5206_5208_synthesis.ps}
%\caption{The Cr\,{\sc i} 5204 \AA\ region showing a comparison of the observed spectrum vs. the best-fitting synthetic spectrum. The macroturbulent velocity used in the synthesis is 20 km
%s$^{-1}$.}
%      \label{f_cr_synthesis}
%       \end{figure}

\noindent {\bf Mn:} The $gf$-values for Mn\,{\sc i} lines are taken from Blackwell-Whitehead
\& Bergemann (2007 when available or otherwise from the NIST database.
Hyperfine structure was considered for all lines with data taken from Kurucz\footnote{\it
http://kurucz.harvard.edu}, as discussed by Prochaska \& McWilliam (2000).
\vskip 0.2 cm

% \begin{figure}
% \centering
% \includegraphics[width=0.99\columnwidth,height=75mm,angle=0]{MnI_4783_synthesis.ps}
%\caption{The Mn\,{\sc i} 4783 \AA\ region showing a comparison of the observed spectrum vs. the best-fitting synthetic spectrum. The macroturbulent velocity used in the synthesis is 20 km
%s$^{-1}$.}
%      \label{f_mn_synthesis}
%       \end{figure}

\noindent {\bf Fe:} The $gf$-values for Fe\,{\sc i} and Fe\,{\sc ii} lines are from Fuhr \& Wiese
(2006). Exclusion of three relatively strong Fe\,{\sc i} lines at 5232 (EW:224 m\AA\ ), 5429 (EW: 220
m\AA\ ), and 5446 \AA\  (EW:206 m\AA\ ) changes the Fe abundance only $-$0.02 dex. Analysis of these lines was discussed in Section  4.1.
\vskip 0.2 cm

\noindent {\bf Co:} The search for Co\,{\sc i} lines drew on the tabulation of $gf$-values
provided by the NIST database. Few Co\,{\sc i} lines are expected to be present: the leading
candidates are lines at 3995.31, 4121.32, and 4118.77 \AA. The 3995\AA\ line, after
allowance for a blending Fe\,{\sc i} line gives the abundance in Table 3. The other two lines are
present but blended.
\vskip 0.2 cm

\noindent {\bf Ni:} The $gf$-values for the Ni\,{\sc i} lines are taken from the
NIST database. A good selection of Ni\,{\sc i} lines is available for an
abundance analysis.
\vskip 0.2 cm

\noindent {\bf Cu:} Copper through Cu\,{\sc i} lines of RMT2 is not detectable in the
spectrum. The strongest line of RMT2 at 5105.54\AA\ gives the upper limit in
Table 3 with the line's  $gf$-value taken from Bielski (1975).
\vskip 0.2 cm

\noindent {\bf Zn:} Zinc is represented by the three Zn\,{\sc i} lines of RMT2 at
4722, 4680, and 4810 \AA.
The $gf$-values are from Bi\'{e}mont \& Godefroid (1980).
\vskip 0.2 cm

\noindent {\bf Sr:} The Sr\,{\sc ii} resonance lines at 4077 \AA\ and 4215 \AA\ are present as
very strong lines and too strong for a reliable abundance determination, i.e., they have EWs of 473 m\AA\ and 408
m\AA\ , respectively.
\vskip 0.2 cm

\begin{figure}
 \centering
 \includegraphics[width=0.99\columnwidth,height=75mm,angle=0]{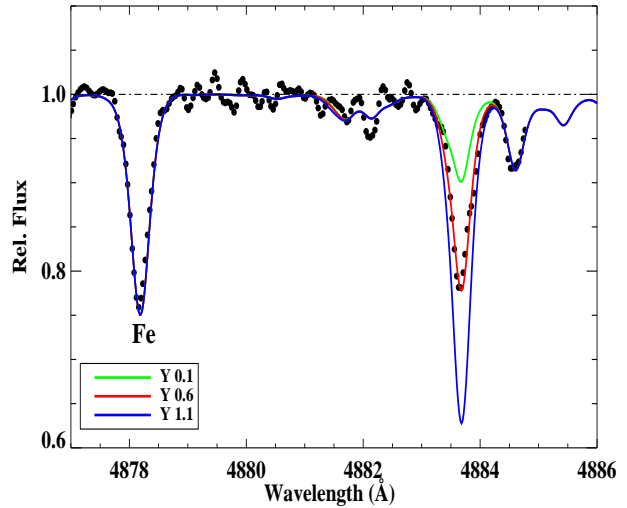}
\caption{Observed and synthetic spectra  for the Y\,{\sc ii} line at 4883 \AA\ .}
     \label{f_YII_synthesis}
 \end{figure}

\noindent {\bf Y:} Selection of Y\,{\sc ii} lines is based on the solar lines judged to be
unblended Y\,{\sc ii} lines in the solar spectrum by Hannaford et al. (1982)  who
provide accurate $gf$-values. Figure~\ref{f_YII_synthesis} shows the best-fitting synthetic spectrum for
the Y\,{\sc ii} line at 4883 \AA\ .
\vskip 0.2 cm

\begin{figure}
 \centering
 \includegraphics[width=0.99\columnwidth,height=75mm,angle=0]{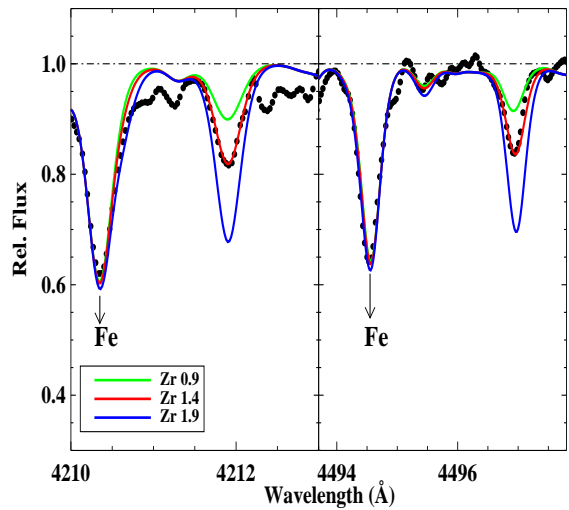}
\caption{Observed and synthetic spectra  for the Zr\,{\sc ii} lines at 4211 \AA\ and 4496 \AA\ .}
     \label{f_zr_synthesis}
 \end{figure}

\noindent {\bf Zr:} Our search for Zr\,{\sc ii} lines drew on the papers by Hannaford et al. (1981)
and Ljung et al. (2006) who
measured accurate laboratory $gf$-values and
conducted an analysis of Zr\,{\sc ii} lines to determine the solar Zr abundance.
Figure~\ref{f_zr_synthesis} shows synthetic spectra fits to two Zr\,{\sc ii} lines.
\vskip 0.2 cm

\noindent {\bf Ba:}  The Ba abundance is based on the 5854\AA\ Ba\,{\sc ii} line, the
weakest line of RMT 2. The $gf$-value  adopted
is the mean of the experimental values from
Gallagher (1967) and Davidson et al. (1992). Hyperfine and isotopic splittings are taken into account
from McWilliam
(1998). The stronger lines from RMT 2 and the resonance lines (RMT 1) give a roughly 0.7 dex
higher Ba abundance and are not considered further.
\vskip 0.2 cm

\noindent {\bf La:} The La\,{\sc ii} line at 5114 \AA\ region was too weak to measure
but used to set the upper limit: $\log\epsilon$(La) $\leq$ 0.31. The $gf$-value for
the line is taken from Lawler et al. (2001a). 
\vskip 0.2 cm

\noindent {\bf Nd:} The absent Nd\,{\sc ii} line at 4303 \AA\ is used to set an upper limit:
$\log\epsilon$(Nd) $\leq$ 0.44. The $gf$-value for the line is taken from 
Den Hartog et al. (2003).
\vskip 0.2 cm

\noindent {\bf Eu:} The Eu\,{\sc ii} resonance lines are 4129\AA\ and 4205\AA\ were
searched for in the spectra. The $gf$-values,
hyperfine and isotopic structure are taken from  Lawler et al. (2001b). Spectrum
synthesis of the 4129\AA\ line gives the abundance in Table 5. The 4205\AA\ line is too
seriously blended to yield a useful a Eu abundance.

 \begin{table}
    \caption[]{Sensitivity of the derived abundances to the uncertainties in the model atmosphere parameters for the $T_{\rm eff}=6500$ K, $\log$ g = 0.5 model.}
       \label{}
   $$
       \begin{array}{@{}l||rrcc}
          \hline
          \hline
         &                 & \Delta log\,\epsilon&     &             \\
\cline{2-5}
$Species$         &  \Delta T_{\rm eff} & \Delta log\,g & \Delta\,\xi  &\Delta\,\xi   \\
\cline{2-5}

  &    $+150$       &  $+0.5$      &  $+1.0$     & $-1.0$ \\
  &       ($K$)     &   ($cgs$)    & ($km$\,s^{-1})& ($km$\,s^{-1})\\
          \hline
C$\,{\sc i}$      &+0.03 &+0.05 &-0.05 &+0.07  \\
N$\,{\sc i}$      &-0.06 &+0.14 &-0.03 &+0.04  \\
O$\,{\sc i}$      &-0.04 &+0.09 &-0.01 &+0.01  \\
Na$\,{\sc i}$     &+0.09 &-0.07 &-0.01 &+0.01  \\
Mg$\,{\sc i}$     &+0.10 &-0.07 &-0.02 &+0.03  \\
Mg$\,{\sc ii}$    &-0.04 &+0.16 &-0.11 &+0.18  \\
Si$\,{\sc i}$     &+0.10 &-0.06 &-0.01 &+0.03  \\
Si$\,{\sc ii}$    &-0.06 &+0.12 &-0.04 &+0.04  \\
S $\,{\sc i}$     &+0.08 &-0.01 &-0.02 &+0.02  \\
Ca$\,{\sc i}$     &+0.10 &-0.07 &-0.04 &+0.05  \\
Ca$\,{\sc ii}$    &-0.01 &+0.12 &-0.00 &+0.01  \\
Sc$\,{\sc ii}$    &+0.08 &+0.13 &-0.02 &+0.04  \\
Ti$\,{\sc i}$     &+0.15 &-0.06 &-0.01 &+0.02  \\
Ti$\,{\sc ii}$    &+0.08 &+0.13 &-0.03 &+0.06  \\
V $\,{\sc ii}$    &+0.05 &+0.11 &-0.07 &+0.06  \\
Cr$\,{\sc i}$     &+0.14 &-0.06 &-0.02 &+0.03  \\
Cr$\,{\sc ii}$    &+0.04 &+0.12 &-0.02 &+0.03  \\
Mn$\,{\sc i}$     &+0.14 &-0.06 &-0.01 &+0.03  \\
Fe$\,{\sc i}$     &+0.11 &-0.08 &-0.03 &+0.02  \\
Fe$\,{\sc ii}$    &+0.05 &+0.13 &-0.03 &+0.03  \\
Co$\,{\sc i}$     &+0.15 &-0.06 &-0.09 &+0.13  \\
Ni$\,{\sc i}$     &+0.12 &-0.06 &-0.02 &+0.03  \\
Cu$\,{\sc i}$     &+0.16 &-0.05 &-0.00 &+0.01  \\
Zn$\,{\sc i}$     &+0.13 &-0.05 &-0.01 &+0.03  \\
%Sr$\,{\sc ii}$    &+0.14 &+0.07 &-0.28 &+0.15  \\
Y $\,{\sc ii}$    &+0.09 &+0.12 &-0.04 &+0.06  \\
Zr$\,{\sc ii}$    &+0.09 &+0.13 &-0.02 &+0.02  \\
Ba$\,{\sc ii}$    &+0.17 &+0.01 &-0.04 &+0.06  \\
La$\,{\sc ii}$    &+0.13 &+0.10 &-0.00 &+0.01  \\
Nd$\,{\sc ii}$    &+0.15 &+0.06 &-0.02 &+0.02  \\
Eu$\,{\sc ii}$    &+0.13 &+0.09 &-0.08 &+0.15  \\
          \hline
          \hline
       \end{array}
   $$
 \end{table}

\noindent Mean abundances are summarized in Table 5. Absolute uncertainties for the abundances arising from
uncertainties of the atmospheric parameters $T_{\rm eff}$, $\log\,g$, and $\xi$ are summarized in Table 6 for
changes with respect to the model of +150 K, +0.5 cm s$^{-2}$, and $\pm$1.0 km s$^{-1}$ for representative lines
(i.e., the Co entries are based on the EW upper limits for the 3995 m\AA\ and 4121 m\AA\ lines). From the
uncertainties listed in Table 6, we find the total absolute uncertainty to be ranging from 0.08 for C\,{\sc i}
to 0.19 for Mg\,{\sc ii} by taking the
square root of the sum of the square of individual errors (for each species) associated with uncertainties in
temperature, gravity, and microturbulent velocity. In light of the line-to-line scatter of abundances, the
absolute uncertainties, and, more importantly, the probability  that the star fails to recognize the suite of
assumptions behind the model atmosphere and the analysis of the lines, specification of abundances to $\pm$0.01
dex is surely astrophysical hubris. Thus, we cite values to 0.1 dex in discussion of the abundances.

\noindent  In the Introduction, we referred to two results of the abundance
analysis by Klochkova (1995), the only previously reported analysis for IRAS 18095+2704. The two results referred to
are: (i) large differences
between the abundance from neutral and ionized lines of several elements, and (ii) overabundances (relative
to iron) of lanthanides. Before discussing the star as a representative of post-AGB stars, we comment
on (i) in light of our analysis. Point (ii) is more properly considered in discussing the star
as a member of the post-AGB class.
\\
\noindent Previous results for abundance differences between neutral and ionized lines
were huge in several cases: i.e., 1.6 for Ti, 1.4 for V, 1.0 for Cr, and 2.2 for Y but 0.0 dex
for Fe due to the fact that it was the condition imposed in determining the surface gravity.
Mg\,{\sc i} and Mg\,{\sc ii} also gave consistent results ($-0.1$ dex). Our results provide
much smaller values, i.e., difference in abundances [X/Fe] derived from neutral and
first-ionized lines of several elements are: $-0.1$ (Ti), and 0.2 (Cr) with again 0.0 dex for
Fe. We were unable to detect V\,{\sc i} lines. For silicon, the
disparity between abundances from Si\,{\sc i} lines and the single weak line of Si\,{\sc ii}
from RMT 5 is $-0.5$ dex. For Ca, it is $+0.1$ dex. We attribute elimination of the earlier large
differences between the abundance from neutral and ionized lines to application of stringent
requirements for the identification of weak lines of neutral species. What about
non-LTE effects? The fact that
the strongest lines generally give abundances that are too large according to a considerations
of astrophysical plausibility is likely due to non-LTE effects: examples include the O\,{\sc
i} 7770\AA\ triplet, the Si\,{\sc ii} 6347\AA\ and 6371\AA\ lines and the resonance lines of
Sr\,{\sc ii} and Ba\,{\sc ii}.

\section{IRAS 18095+2704: abundance anomalies and relatives}

\noindent While IRAS 18095+2704 may be a representative post-AGB star, it is not a textbook example of a star in
this evolutionary state where the AGB star experienced thermal pulses that enrich the envelope in products of
He-shell burning, i.e., carbon directly from the $3\alpha$-process and heavy elements from the $s$-process
probably powered by the $^{13}$C($\alpha, n)^{16}$O reaction. Then, after a series of thermal pulses that
provide for a C-rich star with enhancements of $s$-process nuclides, the star experienced severe mass loss and
became a post-AGB star with effective temperature increasing over a few thousand years. AGB stars of
intermediate mass may experience H-burning at the base of the deep convective envelope (`hot bottom burning') 
such that the C-rich  envelope (with $s$-process enhancements) is reconverted to an O-rich but N-rich envelope
before mass loss sets it on its post-AGB course. In several ways, IRAS 18095+2704 fails to fit these scenarios.
\vskip 0.2 cm

\noindent Our assessment of the evolutionary status of IRAS 18095+2704 involves (i) identifying 
abundance anomalies with respect to the composition of an unevolved or much less evolved
star of the same iron abundance of [Fe/H] $= -0.9$ as obtained from samples of halo and
(thick) disk stars (see, for example, Reddy et al. 2006), and (ii) identifying highly evolved and
likely also post-AGB stars with compositions similar to those of IRAS 18095+2704. 
\vskip 0.2 cm

\noindent We begin by considering those
elements whose atmospheric abundance is expected to be unaffected during the course of stellar evolution.
Such  elements are represented in Table 3 and Table 4 by the run from Mg to Zn, although Mg and Al are possibly
affected by internal nucleosynthesis. In the Mg to Zn sequence, the so-called $\alpha$-elements are
expected to be overabundant relative to Fe by comparable amounts, say [$\alpha$/Fe] $\simeq +0.3$,
provided that the star is truly metal-poor. 
For IRAS 18095+2704, the
$\alpha$-elements have [$\alpha$/Fe] = +0.5 (Mg), +0.6 (Si), +0.8 (S), 0.0 (Ca), and 0.0 (Ti). 
At first glance, these values seem at odds with the roughly equal ratios for these elements
according to studies of disk and halo stars; Mg, Si, and S are more enhanced and Ca and Ti
are less enhanced relative to Fe than in the reference stars with [Fe/H] $\simeq -0.9$. 
One should not overinterpret this result by overlooking potential systematic errors.
It may be significant that [$\alpha$/Fe] is roughly correlated with the atom's ionization
potential and, if this is the case, it may be a signature of systematic errors in either
our analysis and/or in the analyses of the very different stars including main sequence
stars providing the reference [$\alpha$/Fe] values. Across the Fe-group, say Sc to Zn, the ratios [X/Fe] for IRAS 18095+2704 are normal to within
the observational uncertainties, that is [X/Fe] falls within the range $\pm0.2$ dex. Just possibly,
Mn appears overabundant at [Mn/Fe]$=-0.1$ where a value of $-0.4$ is found for disk and halo
stars of [Fe/H] $=-0.9$ (Reddy et al. 2006). 
\vskip 0.2 cm

\noindent Among the suite of elements thought to be immune to internal nucleosynthesis is Al for which we find
[Al/Fe] to be -0.7. This is far from the expected value of
about 0.0. Perhaps, the degree of ionization of Al departs from LTE values and Al atoms
are over-ionized relative to LTE. In some post-AGB stars, elements like Al which condense
readily into grains are underabundant in the stellar atmosphere 
(see, for example, Giridhar et al. 2005), and, then, the elemental underabundances correlate
fairly well with the condensation temperature. This is not the case for IRAS 18095+2704. For example,
Zn with its 726 K condensation temperature (Lodders 2003) is underabundant by [Zn/H] $\simeq -0.7$ but
S at 680K and Na at 958K are barely underabundant, say [S/H] $\simeq$ [Na/H] $\simeq -0.1$.  
In a few post-AGB stars, the underabundances correlate with the neutral atom's ionization potential
(Rao \& Reddy 2005) rather than the condensation temperature
but again IRAS 18095+2704 does not fit this pattern.
\vskip 0.2 cm

\noindent It remains to discuss light elements C, N, O, and Na and the heavy elements Y, Zr, Ba,
La, and Nd which are candidates for abundance adjustments by nucleosynthesis in the course of
stellar evolution. Clearly, the atmosphere is O-rich (C/O $\simeq$ 0.25 by number). Yet, the C
abundance suggests some C enrichment because [C/Fe] $< 0$ is expected following the first
dredge-up but [C/Fe] = +0.4 is found here. The O enhancement ([O/Fe] $=+0.8$) is greater than
reported for normal stars but the excess seems to mirror that found for S, another element of
higher than average ionization potential for an $\alpha$-element and the ratio [O/S] $\simeq$ 0
is similar to that for normal stars. The upper limit to the N abundance is approximately consistent that expected from the
first dredge-up. In short, the C, N, and O abundances suggest mild C-enrichment on the AGB
following dilution of C and enhancement of N during the first dredge-up as the star became a red
giant following the main sequence (see, for example, Iben \& Renzini 1983). Textbook
enrichment of C by thermal pulses with subsequent C-destruction by hot bottom burning is ruled
out by the lack of appreciable N enrichment. Sodium stands apart from this picture: [Na/Fe] $=
+0.8$ is neither a ratio found among normal stars nor easily accounted for on nucleosynthetic
grounds. The Na
abundance appears to be affected by non-LTE effects. Non-LTE analysis of Na I lines by Takeda \&
Takada-Hidai (1994) indicates the non-LTE corrections ($\Delta log\epsilon(X)$)\footnote{$\Delta
log\epsilon(X)$=$\log\epsilon$(X)$_{NLTE}$ - $\log\epsilon$(X)$_{LTE}$} in Na\,{\sc i} lines at
4979, 5683, 5688, and 8195 \AA\ is -0.06, -0.15, -0.20 and -0.94 dex respectively at temperature
6000 K and gravity 0.5 dex.  
 
\vskip 0.2 cm

\noindent Operation of the $s$-process would be expected to lead to overabundances of Y and Zr
but underabundances (relative to Fe) are found. At [Fe/H] $\simeq -1$, Y and Zr underabundances
are not found among normal stars. Thus, the Y and Zr abundances for IRAS 18095+2704 present a
puzzle. It is not clear if the same puzzle is provided by the heavier elements Ba, La, and Nd for
which the $s$-process would provide some enrichment. Barium is nominally consistent with the Y
and Zr in suggesting a slight underabundance ([Ba/Fe] $=-0.2$). Our search for La and Nd proved
unsuccessful and the upper limits [La/Fe] $\simeq$ [Nd/Fe] $\simeq 0$ are consistent with results
for normal stars and also with the Y, Zr, and Ba abundances for IRAS 18095+2704. The $r$-process Eu abundance
is that expected for a normal star; any $s$-process contribution to Eu is expected to be very
slight even had the $s$-process operated in IRAS 18095+2704 (Sneden et al. 2010). In summary, Y
and Zr  underabundances represent a puzzle. One is struck by the fact that Y and Zr with Al have
condensation temperatures among the highest of the elements in Table 4. All are underabundant
relative to expectation. We noted above that the abundances do not always correlate well
with condensation temperature. But elements with condensation
temperatures hotter than 1550K generally appear among the most underabundant - the set includes
Al, Ca, Sc, Ti, Y, Zr, and Ba. (La and Nd also fall in the set but upper limits to their
abundances restricts their relevance here.) At [Fe/H] $\leq -0.3$, the Ca and Ti abundance should
be judged with respect to their overabundance relative to Fe by about 0.3 dex in normal stars,
i.e., add about $-0.3$ dex to the entries for [X/Fe] in Table 5 when judging abundance anomalies
according to condensation temperature. Then, the [X/Fe] for Al, Ca, Ti, Y, Zr, and Ba are quite
similar for these elements of a similar condensation temperature. Scandium appears to be mildly
overabundant with respect to the speculation that elements of the highest condensation
temperature (T$_C > 1500$ K) are underabundant in IRAS 18095+2704.

\begin{figure}
 \centering
 \includegraphics[width=87mm,height=85mm,angle=0]{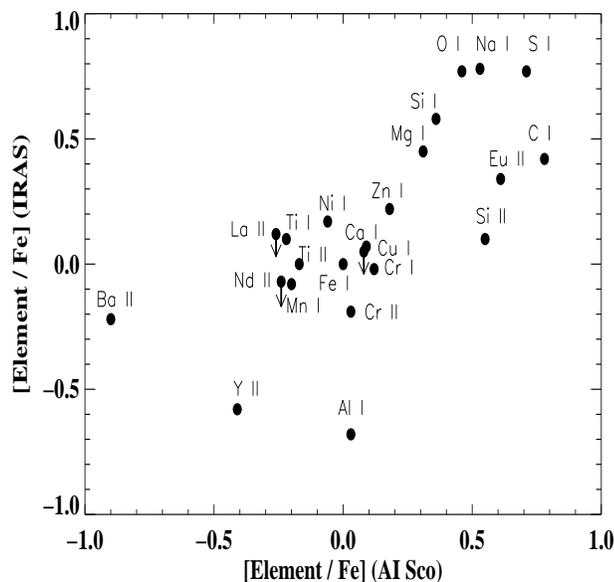}
\caption{[Element/Fe] of AI Sco vs. [Element/Fe] of IRAS 18095+2704. The arrows
indicate to upper limits.}
      \label{f_AI_SCO}
 \end{figure}

In spite of the above remarks about abundances and condensation temperatures, similarities with the compositions of some
RV Tauri variables are present. Among RV Tauri variables and the presumably closely related W Vir variables (Maas et al.
2007), the correlation between abundance anomalies and condensation temperature runs from strong to weak. IRAS 18095+2704
definitely falls among the latter group. There is a fair correspondence between the composition of IRAS 1805+2704 and the
RV Tauri variable AI Sco (Giridhar et al. 2005) (Figure~\ref{f_AI_SCO}).\footnote{There is little resemblance to the
composition of EQ Cas and CE Vir for which abundance anomalies correlate well with the atom's ionization potential.}
 In
part, Figure~\ref{f_AI_SCO} may reflect the fact that systematic errors (e.g., non-LTE effects) are of similar magnitude
for the two stars with similar atmospheric parameters: ($T_{\rm eff}$, $\log\,g$, $[Fe/H]$)=(6500, +0.5, $-$0.9) for IRAS
18095+2704 and (5300, 0.25, -0.7) for AI Sco. Stars exhibiting a striking correlation involving condensation temperature
seem certain to be binaries with a circumbinary disk providing infall of gas but not dust onto the star responsible for
the anomalies (Van Winckel 2003). For those stars (e.g., AI Sco) with hints of a correlation involving the condensation
temperature, their binary status is unknown.  Certainly, the lack of a strong radial velocity variation for IRAS
18095+2704 may suggest that it is a single star. But, perhaps, a wind off the star, as may be suggested by the strong blue
asymmetry for strong lines, provides the site for dust-gas separation. In summary, we suppose that IRAS 18095+2704 may be
related to a RV Tauri variable, probably one that has evolved to hotter temperatures beyond the instability strip. 
\vskip 0.2 cm

\section{Concluding remarks}

Abundance analysis of IRAS 18095+2704 classified as a proto-planetary nebula by Hrivnak et al. (1988)  
suggests the star left the AGB before thermal pulses had the opportunity to enrich the
atmosphere in the principal products from He-shell burning (C and $s$-process nuclides). The star
is not exceptional in this regard - see, for example, the abundance analyses of post-AGB stars
reviewed by Van Winckel (2003). If the star's Fe abundance is adopted as a reference, the
composition of IRAS 18095+2704 is normal except for an underabundance of
Al, Y and Zr. A speculation was offered that elements including Al, Y, and Zr having a condensation temperature hotter
than 1500 K are  underabundant. This suggests that the atmosphere is depleted in those elements that condense
most readily into dust grains. Perhaps, the wind which is suggested by the pronounced blue
asymmetry to the lines of strong lines removes grains selectively. 

Although a variety of spectroscopic indicators provide a consistent set of atmospheric
parameters, the high-resolution optical spectra offer evidence that IRAS 18095+2704's atmosphere
is only an approximation to the classical atmosphere and standard LTE analysis techniques used to
derive the abundances. The line profiles require macroturbulent velocities that are
supersonic. Strong lines suggest by the presence of blue asymmetries that a wind is driven
off the star. Refinement of the abundance analysis may primarily depend on developing an
improved understanding of the physics of these dilute extended stellar photospheres.

\section{Acknowledgments}

This research has been supported in part by the grant F-634 to DLL from the Robert A. Welch Foundation of
Houston, Texas. VGK \& NST acknowledge support from the Russian Foundation for  Basic Research (project No.\,08--02--00072\,a).


\begin{thebibliography}{99}


\bibitem[\protect\citeauthoryear{Aldenius et al.}{2009}]{aldenius09} 
Aldenius M., Lundberg H., Blackwell-Whitehead R., 2009, A\&A, 502, 989

\bibitem[\protect\citeauthoryear{Asplund}{2000}]{asplund00} 
Asplund M., 2000, A\&A, 359, 755 

\bibitem[\protect\citeauthoryear{Asplund et al.}{2009}]{asplund09} 
Asplund M., Grevesse N., Sauval A.~J., Scott P., 2009, ARA\&A, 47, 481

\bibitem[\protect\citeauthoryear{Becker et al.}{1980}]{becker80} 
Becker U., Zimmermann P., Holweger H., 1980, Geochimica et Cosmochimica Acta, 44, 2145 

\bibitem[\protect\citeauthoryear{Bielski}{1975}]{bielski75}
Bielski A., 1975, JQSRT, 15, 463

\bibitem[\protect\citeauthoryear{Biemont \& Godefroid}{1980}]{biemont80} 
Bi\'{e}mont E., Godefroid M., 1980, A\&A, 84, 361 

\bibitem[\protect\citeauthoryear{Biemont et al.}{1981}]{biemont81} 
Bi\'{e}mont E., Grevesse N., Hannaford P., Lowe R.~M., 1981, ApJ, 248, 867 

\bibitem[\protect\citeauthoryear{Biemont et al.}{1989}]{biemont89} 
Bi\'{e}mont E., Grevesse N., Faires L.~M., Marsden G., Lawler J.~E., 1989, A\&A, 209, 391

\bibitem[\protect\citeauthoryear{Blackwell-Whitehead \& Bergemann}{2007}]{blackwell07} 
Blackwell-Whitehead R., Bergemann M., 2007, A\&A, 472, L43

\bibitem[\protect\citeauthoryear{Davidson et al.}{1992}]{1992A&A...255..457D} 
Davidson M.~D., Snoek L.~C., Volten H., Doenszelmann A, 1992, A\&A, 255, 457

\bibitem[\protect\citeauthoryear{Den Hartog et al.}{2003}]{denhartog03} 
Den Hartog E.~A., Lawler J.~E., Sneden C., Cowan J.~J., 2003, ApJS, 148, 543 

\bibitem[\protect\citeauthoryear{Eder, Lewis \& Terzian}{1988}]{Lewis88} 
Eder J., Lewis B. M., Terzian Y., 1988, ApJS, 66, 183

\bibitem[\protect\citeauthoryear{F\"{u}hr \& Wiese}{2006}]{fuhr2006} 
F\"{u}hr J. R., Wiese W. L., 2006, J. Phys. Chem. Ref. Data, 35,  1669

\bibitem[\protect\citeauthoryear{Gallagher}{1967}]{gallagher67} 
Gallagher A., 1967, Phys. Rev., 157, 24

\bibitem[\protect\citeauthoryear{Garz}{1973}]{garz73} 
Garz T., 1973, A\&A, 26, 471 

\bibitem[\protect\citeauthoryear{Giridhar et al.}{2005}]{giridhar05} 
Giridhar S., Lambert D.~L., Reddy B.~E., Gonzalez G., Yong D., 2005, ApJ, 627, 432 

\bibitem[\protect\citeauthoryear{Gledhill et al.}{2001}]{gled2001} 
Gledhill T.~M., Chrysostomou A., Hough J.~H., Yates J.~A., 2001, MNRAS, 322, 321 


\bibitem[\protect\citeauthoryear{Hannaford et al.}{1982}]{hanna82} 
Hannaford P., Lowe R.~M., Grevesse N., Bi\'{e}mont E., 1982, ApJ, 261, 736

\bibitem[\protect\citeauthoryear{Horne}{1986}]{horne86} 
Horne K.~D., 1986, PASP, 98, 609

\bibitem[\protect\citeauthoryear{Howarth \& Phillips}{1986}]{howarth86} 
Howarth I.~D., Phillips A.~P., 1986, MNRAS, 222, 809 

\bibitem[\protect\citeauthoryear{Howarth et al.}{1998}]{howarth98} 
Howarth I.~D., Murray J., Mills D., Berry D.~S., 1998, Starlink User Note 50

\bibitem[\protect\citeauthoryear{Hrivnak, Kwok, \& Volk}{1987}]{hrivnak87} 
Hrivnak B.~J., Kwok S., Volk K.~M., 1987, BAAS, 19, 1091

\bibitem[\protect\citeauthoryear{Hrivnak, Kwok, \& Volk}{1988}]{hrivnak88} 
Hrivnak B.~J., Kwok S., Volk K.~M., 1988, ApJ, 331, 832

\bibitem[\protect\citeauthoryear{Iben \& Renzini}{1983}]{Iben83} 
Iben I.~Jr., Renzini A., 1983, ARA\&A, 21, 271

\bibitem[\protect\citeauthoryear{Kelleher \& Podobedova}{2008}]{Kelleher08} 
Kelleher D.~E., Podobedova L.~I., 2008, J. Phys. Chem. Ref. Data, 37, 1285

\bibitem[\protect\citeauthoryear{Klochkova}{1995}]{kloc95} 
Klochkova V.~G., 1995, MNRAS, 272, 710

\bibitem[\protect\citeauthoryear{Kurucz}{1993}]{kurucz93} 
Kurucz R.~L., 1993, Kurucz CDROM Vol 18 (Cambridge: Smithsonian Astrophysical Observatory)

\bibitem[\protect\citeauthoryear{Lawler \& Dakin}{1989}]{lawler89} 
Lawler J.~E., Dakin J.~T., 1989, JOSA, B6, 1457

\bibitem[\protect\citeauthoryear{Lawler et al.}{2001}]{lawler01a} 
Lawler J.~E., Bonvallet G., Sneden C., 2001a, ApJ, 556, 452

\bibitem[\protect\citeauthoryear{Lawler et al.}{2001}]{lawler01b} 
Lawler J.~E., Wickliffe M.~E., Den Hartog E.~A., 2001b, ApJ, 563, 1075

\bibitem[\protect\citeauthoryear{Lewis, Eder \& Terzian}{1985}]{Lewis85} 
Lewis B.~M., Eder J., Terzian Y., 1985, Nature, 313, 200

\bibitem[\protect\citeauthoryear{Ljung et al.}{2006}]{Ljung06} 
Ljung G., Nilsson H., Asplund M., Johansson S., 2006, A\&A, 456, 1181

\bibitem[\protect\citeauthoryear{Lodders}{1985}]{Lodders85}
Lodders K., 2003, ApJ, 591, 1220

\bibitem[\protect\citeauthoryear{Maas et al.}{2007}]{maas07} 
Maas T., Giridhar S., Lambert D.~L., 2007, ApJ, 666, 378 

\bibitem[\protect\citeauthoryear{McWilliam}{1998}]{McWilliam98} 
McWilliam A., 1998, AJ, 115, 1640

\bibitem[\protect\citeauthoryear{Mills \& Webb}{1994}]{mill94}
Mills D., Webb J., 1994, Rutherford Appleton Laboratory, SUN 152.1

\bibitem[\protect\citeauthoryear{Moore}{1945}]{moore45} Moore C.~E., 1945, "A
Multiplet Table of Astrophysical Interest", Princeton Obs. Contr. No. 20
(reprinted 1959, Nat. Bur. Stand. Technical Note 36)

\bibitem[\protect\citeauthoryear{Nilsson et al.}{2006}]{nilsson06} 
Nilsson H., Ljung G., Lundberg H., Nielsen K.~E., 2006, A\&A, 445, 1165

\bibitem[\protect\citeauthoryear{Osterbrock \& Ferland}{2006}]{osterbrok06} 
Osterbrock D.~E., Ferland G.~J., 2006, Astrophysics of gaseous nebulae and active galactic nuclei, Sausalito, CA: University Science Books  

\bibitem[Panchuk et al., 2007]{nes} 
Panchuk V.~E., Klochkova V.~G., Yushkin M.~V., Najdenov I.~D., {\it Proceedings of the Joint Discussion No.\,4 during the IAU General
    Assembly of 2006} Ed. by A.~I.~Gomez de Castro and M.~A.~Barstow, (Editorial Complutense, Madrid, 2007), p.179.

\bibitem[Pickering et al. 2001]{pickering01} 
Pickering J.~C., Thorne A.~P., Perez R., 2001, ApJS, 132, 403

\bibitem[\protect\citeauthoryear{Pickering et al.}{2002}]{pickering02} 
Pickering J.~C., Thorne A.~P., Perez R., 2002, ApJS, 138, 247 

\bibitem[\protect\citeauthoryear{Podobedova et al.}{2009}]{podobedova09}
Podobedova L.~I., Kelleher D.~E., Wiese W.~L., 2009, J. Phys. Chem. Ref. Data 38, 171

\bibitem[\protect\citeauthoryear{Prochaska \& McWilliam}{2000}]{prochaska00} 
Prochaska J.~X., McWilliam A., 2000, ApJ, 537, L57

\bibitem[\protect\citeauthoryear{Rao \& Reddy}{2005}]{rao2005} 
Rao N.~K., Reddy B.~E.\ 2005, MNRAS, 357, 235

\bibitem[\protect\citeauthoryear{Reddy et al.}{2006}]{reddy06} 
Reddy B.~E., Lambert D.~L., Allende Prieto C., 2006, MNRAS, 367, 1329 

\bibitem[\protect\citeauthoryear{\c{S}ahin}{2008}]{sahin08} 
\c{S}ahin T., 2008, Ph.D.~Thesis, Queen's University Belfast

\bibitem[\protect\citeauthoryear{Sch\qo nberner}{1983}]{sc83} 
Sch\qo nberner D., 1983, ApJ, 272, 708

\bibitem[\protect\citeauthoryear{Shi et al.}{2008}]{shi08} 
Shi J.~R., Gehren T., Butler K., Mashonkina L.~I., Zhao G., 2008, A\&A, 486, 303 

\bibitem[\protect\citeauthoryear{Sneden}{2002}]{sneden02} 
Sneden C., 2002, {\sc MOOG} An LTE Stellar Line Analysis Program

\bibitem[\protect\citeauthoryear{Sneden et al.}{2010}]{sneden10} 
Sneden C., Cowan J.~J., Gallino R., 2010, IAU Symposium, 265, 46 

\bibitem[\protect\citeauthoryear{Sobeck et al.}{2007}]{sobeck07} 
Sobeck J.~S., Lawler J.~E., Sneden C., 2007, ApJ, 667, 1267

\bibitem[\protect\citeauthoryear{Takeda \& Takada-Hidai}{1994}]{takeda94} 
Takeda Y., Takada-Hidai M., 1994, PASJ, 46, 395

\bibitem[\protect\citeauthoryear{Tamura et al.}{1993}]{tamura93} 
Tamura S., Takeuti M., Zalewski J., 1993, Ap\&SS, 210, 159 

\bibitem[\protect\citeauthoryear{Tull et al.}{1995}]{tull95} 
Tull R.~G., MacQueen P.~J., Sneden C., Lambert D.~L., 1995, PASP, 107, 251

\bibitem[\protect\citeauthoryear{Van Winckel}{2003}]{winckel03} Van Winckel H., 2003, ARA\&A, 41, 391 

\bibitem[\protect\citeauthoryear{Volk \& Kwok}{1987}]{volk87} 
Volk K., Kwok S., 1987, ApJ, 315, 654

\bibitem[\protect\citeauthoryear{Wiese, F\"{u}hr \& Deter}{1996}]{wiese96}
Wiese W.~L., F\"{u}hr J.~R., Deters T.~M.,1996, J. Phys. Chem. Ref. Data Monograph No. 7

\bibitem[Yushkin \& Klochkova, 2005]{Yushkin} 
Yushkin M.~V., Klochkova V.~G., 2005, Preprint of the Special Astrophysical Observatory No. 206

\end{thebibliography}
\end{document}